\documentclass[lettersize,journal]{IEEEtran}
\usepackage{amsmath,amsfonts,amssymb}
\usepackage{algorithmic}
\usepackage{algorithm}

\usepackage{array}

\usepackage{textcomp}
\usepackage{stfloats}
\usepackage{url}
\usepackage{verbatim}
\usepackage{graphicx}
\usepackage{cite}

\usepackage{subcaption}

\newtheorem{remark}{Remark}

\usepackage{subcaption}
\usepackage{xcolor}
\definecolor{navyblue}{rgb}{0.0, 0.0, 0.5}

\begin{document}

\title{Point Cloud Environment-Based Channel Knowledge Map Construction}
\author{Yancheng Wang,~\IEEEmembership{Graduate Student Member,~IEEE}, Wei Guo,~\IEEEmembership{Member,~IEEE}, Chuan~Huang,~\IEEEmembership{Member,~IEEE}, Guanying Chen,~\IEEEmembership{Member,~IEEE}, Ye Zhang,~\IEEEmembership{Member,~IEEE}, and Shuguang Cui,~\IEEEmembership{Fellow,~IEEE}

\thanks{An earlier version of this paper was presented in part at the IEEE Vehicular Technology Conference, Washington DC, USA, October 2024 \cite{vtc2024}.

Y. Wang is with the Shenzhen Future Network of Intelligence Institute, the School of Science and Engineering, and the Guangdong Provincial Key Laboratory of Future Networks of Intelligence, The Chinese University of Hong Kong, Shenzhen, China (e-mail: yanchengwang@link.cuhk.edu.cn).

W. Guo is with the Department of Electronic and Computer Engineering, The Hong Kong University of Science and Technology, Hong Kong SAR, China (e-mail: eeweiguo@ust.hk).

G. Chen is with the School of Cyber Science and Technology, Sun Yat-sen University, Shenzhen, China (e-mail: chenguanying@mail.sysu.edu.cn).

Y. Zhang is with the School of Computer Science, Beijing Information Science and Technology University, Beijing, China (e-mail: yezhang@bistu.edu.cn).

C. Huang and S. Cui are with the School of Science and Engineering, the Shenzhen Future Network of Intelligence Institute, and the Guangdong Provincial Key Laboratory of Future Networks of Intelligence, The Chinese University of Hong Kong, Shenzhen, China (e-mails: \{huangchuan, shuguangcui\}@cuhk.edu.cn).

}
}

\maketitle

\begin{abstract}
Channel knowledge map (CKM) provides certain levels of channel state information (CSI) for an area of interest, serving as a critical enabler for environment-aware communications by reducing the overhead of frequent CSI acquisition.
However, existing CKM construction schemes adopt over-simplified environment information, which significantly compromises their accuracy. To address this issue, this work proposes a joint model- and data-driven approach to construct CKM by leveraging point cloud environmental data along with a few samples of location-tagged channel information.
First, we propose a novel point selector to identify subsets of point cloud that contain environmental information relevant to multipath channel gains, by constructing a set of co-focal ellipsoids based on different time of arrival (ToAs).
Then, we trained a neural channel gain estimator to learn the mapping between each selected subset and its corresponding channel gain, using a real-world dataset we collected through field measurements, comprising environmental point clouds and corresponding channel data.
Finally, experimental results demonstrate that: For CKM construction of power delay profile (PDP), the proposed method achieves a root mean squared error (RMSE) of 2.95 dB, significantly lower than the 7.32 dB achieved by the conventional ray-tracing method; for CKM construction of received power values, i.e., radio map, it achieves an RMSE of 1.04 dB, surpassing the Kriging interpolation method with an RMSE of 1.68 dB.

\end{abstract}


\section{Introduction}
Channel state information (CSI) is critical for advanced wireless communication technologies, e.g., orthogonal frequency-division multiplexing (OFDM) and multiple-input multiple-output (MIMO) to improve link transmission rate \cite{lopez2020ultra} and mitigate multi-user interference \cite{budhiraja2021interference} in dynamic environment. 
However, with the expectation to support denser users, higher transmission rate, more reliable connections, and lower service delay, next-generation wireless systems require significantly larger number of antennas (i.e., massive MIMO with thousands of antennas), higher frequencies (i.e., mmWave and THz), and wider bandwidths (i.e., several GHz or more).
Under the above scenarios, conventional CSI acquisition methods by utilizing pilots are becoming extremely resource-consuming and impractical due to their substantial overhead and quite limited scalability \cite{wang2023road, han2023towards,zeng2021toward}. 
Moreover, passive devices such as reconfigurable intelligent surfaces (RIS), promising for next-generation wireless systems, lack inherent channel estimation capabilities and entirely rely on active terminals to obtain CSI, further causing significant pilot overhead.
Therefore, a revolutionary method for CSI acquisition is urgently needed to meet the explosive demands in the next-generation wireless networks.

To address the above challenges in CSI acquisition, the concept of channel knowledge map (CKM) was introduced \cite{romero2022radio, wang2020indoor,zeng2021toward,zeng2023tutorial}. 
CKM is essentially a map overlaid with channel parameters across an area of interest (AoI), e.g., received power\cite{romero2022radio,zhangsongyang}, shadowing loss\cite{levie2021radiounet,bakirtzis2022deepray}, multipath information\cite{zeng2021toward}, channel impulse response (CIR)\cite{virk2018site}, and power delay profile (PDP)\cite{virk2018site,koivumaki2021impacts,jarvelainen2016indoor}. With the help of CKM, it is possible to simplify or even eliminate the needs for pilot-based CSI acquisition \cite{zeng2023tutorial}. Moreover, CKM also provides significant value in applications such as base station planning \cite{bethanabhotla2015optimal}, fingerprint-based localization \cite{fingerprint2023gao}, and spectrum allocation \cite{debroy2015spectrum}. However, existing CKM construction methods adopt over-simplified environmental information, resulting in very low construction accuracy. 

\subsection{Related Work} 

Numerous CKM construction methods have been proposed and are broadly categorized into two classes: model- and data-driven methods.

\subsubsection{Model-Driven Methods}
Model-driven methods leverage signal models that capture the underlying electromagnetic (EM) wave propagation characteristics and utilize location-tagged channel information to construct CKM. These methods are further classified into three subcategories: interpolation\cite{chowdappa2018distributed, 42}, virtual-obstacle \cite{romero2018blind,CheEsrGesMit:C17,106}, and ray-tracing methods\cite{da2021ray,he2018design,jung2014ray,_wireless,_ray,krijestorac2021spatial}. 

\emph{Interpolation methods} estimate the channel knowledge by utilizing statistical models \cite{chowdappa2018distributed, 42} to exploit the spatial correlations between the measured and unmeasured channel data, i.e., nearby locations exhibit similar EM wave propagation characteristics for shadowing and scattering. The authors in \cite{chowdappa2018distributed} developed a Kriging interpolation method to construct CKM by exploiting the intrinsic geostatistical dependencies inherent in the measured channel data. Moreover, CKM exhibits an interesting low-rank property over the AoI, since the corresponding pathloss gradually changes with distance \cite{42}. By exploiting this property, the authors in \cite{42} demonstrated that CKM can be well-approximated by a low-rank matrix and proposed a singular value thresholding (SVT) algorithm for CKM construction. 
In summary, the above interpolation methods perform well in relatively simple environments, e.g., rural areas or regions with few obstacles. However, in complex urban environments, the presence of numerous buildings and obstacles introduces substantial shadowing and reflection, leading to non-smooth variations in the CKM and significantly degrading the performance of interpolation methods\cite{levie2021radiounet}.

\emph{Virtual-obstacle methods} simultaneously construct a CKM and a virtual obstacle map, with the latter capturing the shadowing effects induced by various obstacles. The authors in \cite{romero2018blind} proposed a tomographic framework that constructs a spatial loss field (SLF) by discretizing the AoI into uniformly spaced pixels and inferring the per-pixel shadowing loss from the measured data. 
Similarly, the authors in \cite{CheEsrGesMit:C17,106} proposed a hidden multi-class virtual obstacle framework that constructs the CKM by optimizing the heights and classes of virtual obstacles within the environment to align with the measured channel data. Although these methods effectively captured shadowing effects, they still face challenges in handling more complicated propagation phenomena, such as reflection and scattering, due to their inability to incorporate real-world environment. 

\emph{Ray-tracing methods} construct CKM by utilizing both the environment and EM wave propagation models \cite{da2021ray,he2018design,jung2014ray,_wireless,_ray}. This type of methods treats each propagation path as one ray, emitting multiple rays from the transmitter and tracing their trajectories through a 3D environment. Each path obeys the physical principles of reflection, diffraction, and scattering, and only a subset of these rays ultimately reaches the receiver. In \cite{da2021ray}, the authors constructed a simplified 3D environment model by using box-shaped buildings and cylinder-shaped trees, and then derived the physical models for reflection, shadowing, and scattering by using EM theory \cite{Julius2015em} to determine the propagation direction and attenuation of each path. 
Although ray-tracing methods are theoretically more accurate than interpolation methods, their practical performance is often limited by the over-simplified environment models and inaccurate material parameters.

Recent advancements in 3D environment reconstruction technologies, such as LiDAR \cite{wang2018lidar} and photogrammetry \cite{mikhail2001introduction}, have enabled the creation of environment models in the form of point cloud, where the points represent the external surfaces of environmental objects with centimeter-level accuracy. 
To use point cloud as the environment model in ray-tracing, the authors in \cite{jarvelainen2016indoor} constructed flat edges and surfaces by fitting point cloud as lines and planes, and then used the constructed surfaces as the 3D environment model for ray-tracing, which still cannot fully leverage the detailed information in point cloud.

\subsubsection{Data-Driven Methods} 
Recent advancements in deep learning have introduced new data-driven (i.e., AI-based) approaches for CKM construction \cite{levie2021radiounet, bakirtzis2022deepray, zhangsongyang}.
The authors in \cite{levie2021radiounet} reformulated the CKM construction problem as an image-to-image prediction task by converting the environment information and the transmitter location into two images: a 2D environment map and a transmitter map with a dot indicating its location. Using convolutional neural networks (CNNs), the proposed RadioUNet was trained to predict the CKM image from a ray-tracing simulation dataset that included a variety of environments and transmitter locations. Following RadioUNet's success, subsequent research further advanced this approach. In \cite{bakirtzis2022deepray}, manually labeled building material properties were integrated into the input images to provide richer information for training neural networks, and the authors in \cite{zhangsongyang} explored advanced generative adversarial network (GAN) to improve the CKM construction accuracy. 

Despite the above advancements, current data-driven methods, which were only designed for processing 2D environments, struggle to handle 3D data\cite{bakirtzis2022deepray}, e.g., point cloud or voxel grid. Directly converting 3D data to 2D and applying the current 2D methods leads to a huge loss of essential 3D details and significantly degrades the CKM construction accuracy\cite{da2021ray}. 
Moreover, current data-driven methods only focus on estimating the received signal strength at each location, i.e., radio map construction\cite{levie2021radiounet, bakirtzis2022deepray, zhangsongyang}, while do not account for more precise channel knowledge, e.g., PDP and CIR, which requires multi-path channel models to determine the propagation paths.

\subsection{Our Contributions}

This paper addresses the CKM construction problem by directly incorporating the 3D point cloud environmental information. Specifically, we focuses on estimating two types of channel knowledge: the PDP, which captures the channel gains at different ToAs, and the received power values, which collectively form the radio map.
3D environmental point clouds are employed due to their ability to retain rich geometric and color information, enabling accurate representation of complex environmental structures and capturing key EM wave propagation-related features (e.g., color that implicitly encodes material properties, and normal vector that reflects the orientation of object surface). 

However, harnessing 3D point cloud for CKM construction presents significant challenges. First, point cloud lacks surfaces or meshes to perform conventional ray-tracing, making it challenging to identify a specific subset of point cloud that influences a particular propagation path. 
Moreover, point cloud is unordered data structures, and existing methods fail to utilize this data effectively without converting it to other formats, thereby compromising its precision or details. This work, to the best of our knowledge, presents the first joint model- and data-driven CKM construction method that incorporates 3D environmental information into the construction process. Our contributions are summarized as follows:

\begin {enumerate}

\item We propose a model-driven module, named Point Selector, to identify subsets of point cloud that contain environmental information related to the channel gains at different time of arrival (ToA). We refer to these subsets as interacting objects (IOs) since they actively influence EM wave propagation through reflection, scattering, and shadowing. The proposed model identifies IOs from a temporal perspective by constructing a set of co-focal ellipsoids based on different ToAs, enabling a thorough selection of IOs while preserving geometric details and propagation-related features inherent in point cloud. 

\item We propose a data-driven module, named Neural Channel Gain Estimator, which is trained to directly learn the mapping from the IOs to the channel gains. 
The proposed network is specifically designed to extract the inherent structural information in point cloud that influences the corresponding channel gain. Furthermore, the network exploits critical point cloud features, e.g., the color and normal vectors, that impact EM wave propagation to improve the accuracy of channel gain estimation. 

\item We conduct experiments by using our own field-collected channel PDP and environmental point cloud data, which were systematically acquired through real-world measurement campaigns. Extensive comparison experiments demonstrate that by fully utilizing point cloud environment information, the proposed method achieves a root mean squared error (RMSE) of 2.95 dB for CKM construction of PDP, which outperforms the conventional ray-tracing method with RMSE of 7.32 dB. The proposed method also achieves an RMSE of 1.04 dB for CKM construction of received power values, i.e., radio map construction, which outperforms the Kriging interpolation method with RMSE of 1.68 dB.

\end{enumerate}

The remainder of this paper is organized as follows. Section \ref{System Model} formulates the system model of CKM construction problem. Section \ref{alg} presents the proposed CKM construction method. Section \ref{exp} shows the experimental results for CKM constructions. Section \ref{conclusion} concludes this paper.

\section{System Model}
\label{System Model}

For one given AoI $S \subset \mathbb{R}^{3}$ and one transmitter\footnote{The proposed methods can be easily extended to the scenarios with multiple transmitters within the AoI by either aggregating the received power values or selecting the strongest signal at each location.} with known location $\mathbf x_\text{T} \in S$, our objective is to construct the CKM over $S$, by incorporating the 3D point cloud environmental information of $S$ and a limited set of location-tagged channel data. By using the photogrammetry methods \cite{mikhail2001introduction}, 3D point cloud of the environment is reconstructed from multi-view photographs. 
The obtained point cloud is denoted as $\mathcal{P}_\text{env}=\{(\mathbf p_i,\mathbf s_i)|i=1, \cdots, Q\}$, where $\mathbf p_i \in \mathbb{R}^{3}$ is the coordinate of the $i$-th point, $\mathbf s_i \in \mathbb{R}^{C}$ is the concatenation of  additional features, including color, normal vector, and other auxiliary attributes, with $C$ being the feature dimension, and $Q$ is the number of the points in the cloud. Our study focuses on the quasi-static environments, where major objects, e.g., buildings and walls, in the AoI remain stationary over time, and other minor moving objects such as humans are omitted. 

A multipath channel model with a sampling interval $\Delta t$ \cite{tse2005fundamentals} is considered in this paper, and is completely characterized by its discrete-time CIR $h[n]$, i.e., 
\begin{equation}
h[n] = \sum_{i=0}^{K-1} \alpha_i e^{j\phi_i} \delta[n - i],
\label{cir}
\end{equation}
where $K$ is the length of the discrete-time CIR sequence, $\alpha_i$ is the real-valued channel gain at ToA $t_i=t_0+i \Delta t$ with $t_0$ denoting the propagation time of the line-of-sight (LoS) path, $e^{j\phi_i}$ is the phase at ToA $t_i$ with $j$ being the imaginary unit, and $\delta[n - i]$ is the Kronecker delta function with $\delta[n - i]=1$ for $n = i$ and $\delta[n - i]=0$ for $n \neq i$. 

This paper aims to recover two types of channel knowledge: the received power value $P$ and the PDP $\mathbf h$, which, according to the channel model in \cite{tse2005fundamentals}, are defined as $P = \sum_i \alpha_i^2$\cite{power_strength} and $\mathbf h = [\alpha_0, \alpha_1, \cdots, \alpha_{(K-1)}]\in \mathbb{R}^{K}$, respectively. 
As the phase $e^{j\phi_i}$ is not involved to these two definitions, we focus on estimating the real-valued channel gain $\alpha_i$ and disregard the phase\footnote{It has also been revealed in \cite{da2021ray}, that the phase is influenced by many random factors, e.g., small moving objects, and is generally modeled as random parameters and thus ignored in this paper.}.
In the following section, we will only discuss the PDP for brevity. 

This paper utilizes $L$ observed channel knowledge to construct the CKM, with the observed channel knowledge set denoted as $\{(\mathbf{x}_l^o, \mathbf{h}_l^o)\}_{l=1}^{L}$, where $\mathbf x_{l}^o \in \mathbb{R}^{3}$ is the location of the $l$-th observation, $\mathbf h_{l}^o \in \mathbb{R}^{K}$ is the observed PDP at that location, and $\{\mathbf{x}_1^o, \mathbf{x}_2^o, \ldots, \mathbf{x}_L^o\}$ and $\{\mathbf{h}_1^o, \mathbf{h}_2^o, \ldots, \mathbf{h}_L^o\}$ are the sets of observed locations and corresponding PDPs, respectively. 
The primary objective of the CKM construction is to determine a function
\begin{align}\label{eq:mapping}
\mathbf{h}  =f \left( \mathbf x |  \mathcal{P}_\text{env}, \mathbf x_\text{T},\{(\mathbf{x}_l^o, \mathbf{h}_l^o)\} \right): \mathbf x  \in \mathbb{R}^{3}\mapsto \mathbf h \in \mathbb{R}^{K},
\end{align}
that maps any receiver location within the AoI, i.e., $\mathbf{x} \in S$ and $\mathbf{x} \notin \{\mathbf{x}_1^o, \mathbf{x}_2^o, \ldots, \mathbf{x}_L^o\}$, to its corresponding channel knowledge $\mathbf{h} \in \mathbb{R}^{K}$. This mapping leverages the observed channel knowledge set $\{(\mathbf{x}_l^o, \mathbf{h}_l^o)\}_{l=1}^{L}$, the point cloud $\mathcal{P}_\text{env}$, and the transmitter location $\mathbf{x}_\text{T}$.

\begin{remark}
The CKM construction problem in \eqref{eq:mapping} can be theoretically solved by Maxwell's equations with finite element methods (FEM) \cite{li2011comparative}, by replacing the point cloud $\mathcal{P}_\text{env}$ with a 3D mesh model. However, this method faces significant challenges: (1) The computational complexity of FEM is prohibitive for large spatial areas with dense meshing. (2) The precise material parameters (e.g., permittivity, permeability, and conductivity) for all objects in the environment are unknown or hard to measure, yet they are indispensable for solving Maxwell's equations. Therefore, a joint model- and data-driven method is needed to efficiently approximate EM behavior without requiring precise material parameters or incurring the high computational cost of FEM.
\end{remark}

\section{Proposed CKM Construction Method}
\label{alg}

In this section, we propose a novel method for constructing the CKM by leveraging both the 3D environmental point cloud and the observed channel knowledge. 
To directly utilize 3D environmental point cloud without simplification or transformation, we propose the Point Selector that identifies subsets of point cloud related to channel gains at different ToAs. Additionally, by using the measured channel data, we train a Neural Channel Gain Estimator to learn the mapping from the identified subset of point cloud to the channel gain, which is capable of utilizing the rich information inherent in the environmental point cloud. 

\begin{figure*}[htbp]
\centering

\includegraphics[width=0.8\linewidth]{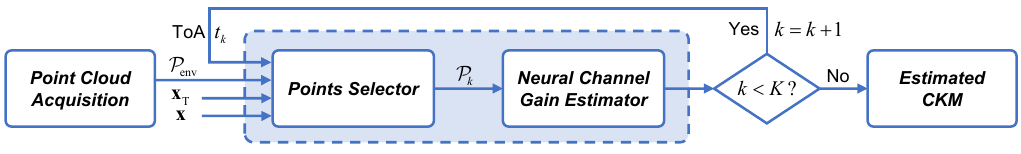}
\caption{Flowchart of the proposed CKM construction method.}
\label{F:framework}
\end{figure*}

As depicted in Fig.~\ref{F:framework}, the proposed CKM construction scheme consists of the following four parts:
\begin{itemize}
    \item Point Cloud Acquisition: The CKM construction scheme starts with point cloud acquisition. Using the photogrammetry method \cite{mikhail2001introduction}, high-precision 3D environment information is obtained in the form of point cloud. The resulting point cloud $\mathcal{P}_\text{env}$, the transmitter location $\mathbf x_\text{T}$, the receiver location $\mathbf x$, and the current ToA $t_k$ are sent to the next module. 

    \item Point Selector: This module identifies relevant points in the point cloud whose corresponding objects affect EM wave propagation. It constructs a sequence of mutually exclusive arriving regions $R_k$, $k=0,\cdots,K-1$, based on different ToA intervals $[t_k,t_{k+1})$, with respect to (w.r.t.) the transmitter and receiver locations $\mathbf x_\text{T}$ and $\mathbf x$, and outputs a subset of point cloud $\mathcal{P}_k$ located in $R_k$.

    \item Neural Channel Gain Estimator: This module estimates the channel gain $\alpha_k$ corresponding to a given ToA and receiver location by leveraging the identified subset of point cloud $\mathcal{P}_k$. 

    \item Estimated CKM: The estimation process iterates through the PDP sequence of length $K$, chronologically combines $K$ channel gain estimations to form the estimated PDP at receiver location $\mathbf x$, and repeats this procedure across the entire AoI to complete the CKM. 
\end{itemize}

In the following, we focus on the two core modules in the proposed CKM construction scheme, i.e., Point Selector and Neural Channel Gain Estimator.

\subsection{Point Selector}

The proposed Point Selector works based on the following two facts: (1) For given ToA $t_k$, there might be multiple propagation paths that simultaneously reach the receiver, all having the same propagation distance $d_k = ct_k$, with $c$ being the speed of light; (2) The sum of the distances from any point on an ellipse to its two foci is a constant. Therefore, we are motivated to use ellipses to identify objects interacting with EM waves arriving at a given ToA. 

\begin{figure}[!t]
\centering
\includegraphics[width=0.75\linewidth]{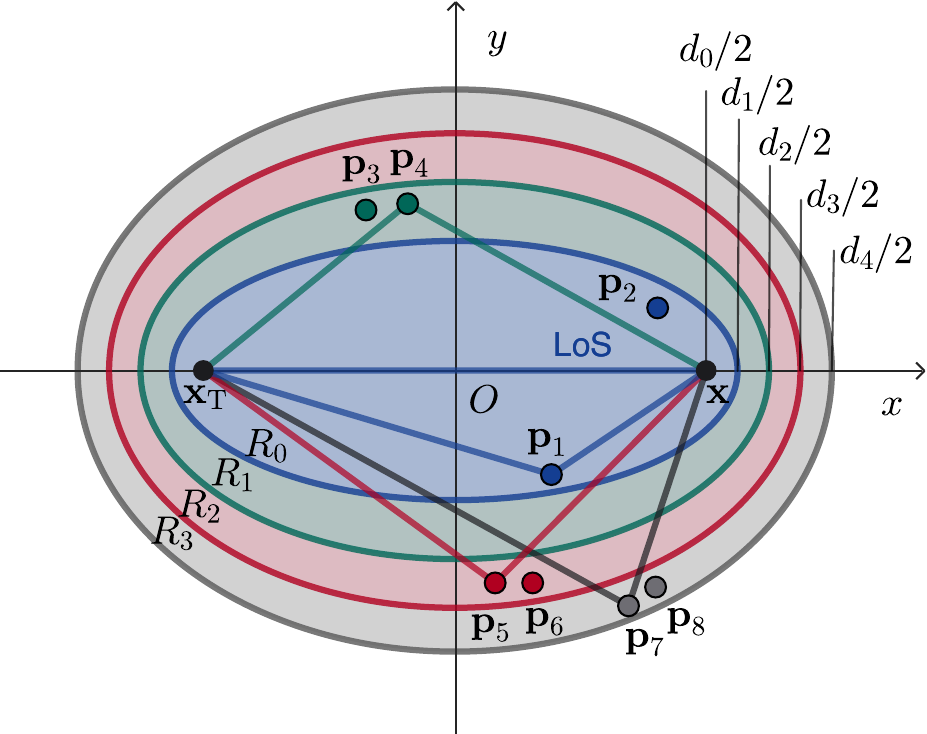}
\caption{2D cross-section view of arrival regions.}
\label{F:fresnelzone}
\end{figure}

Specifically, for 3D scenario, we define an ellipsoid, whose foci are transmitter and receiver locations $\mathbf x_\text{T}$ and $\mathbf x$, such that any point $\mathbf{p}$ on the ellipsoid satisfies $\|\mathbf{p} - \mathbf{x}_\text{T}\| + \|\mathbf{p} - \mathbf{x}\| = d_k$.
For a sequence of ToAs $t_0,\cdots, t_{K-1}$, with $t_k=t_0+k\Delta t$, we construct a sequence of co-focal ellipsoids, as depicted in Fig.~\ref{F:fresnelzone}, segmenting the environment into distinct regions named ``arriving regions''. The first arriving region, denoted by $R_0$, is the space within the innermost ellipsoid. Subsequent arriving regions, denoted by $R_1, R_2, R_3, \cdots$, are the regions between the adjacent ellipsoids, depicted in different colors. Some example points $\mathbf p_1, \cdots ,\mathbf p_8 $ in Fig.~\ref{F:fresnelzone} represent points located in different arriving regions and the solid lines represent the corresponding propagation paths, differentiated with various colors. 
In the following, we detail the construction process of these arriving regions.

\subsubsection{First Arriving Region}
Being the innermost ellipsoid, the first arriving region $R_0$ covers all objects that interact with EM waves arriving within time interval $[t_0,t_1)$, where $t_1 = t_0 + \Delta t$. Due to the finite sampling interval $\Delta t$ of the channel, the receiver accumulates the signal energy over each interval $[t_0, t_{1})$ and assigns it to ToA $t_0$. Thus, all EM waves arriving within this interval are treated as arriving at $t_0$.

To simplify, we consider a 3D Cartesian coordinate system and place transmitter $\mathbf x_\text{T}$ and receiver $\mathbf x$ symmetrically along the x-axis at $(-d_0/2,0,0)$ and $(d_0/2,0,0)$, respectively, where $d_0 = c t_0$ is length of the shortest path between them. All points that interact with the arriving EM waves during time interval $[t_0, t_1)$ are located inside an ellipsoid, whose size is determined by the propagation distance $d_1= ct_1$. 
Hence, the first arriving region $R_0$ is defined as
\begin{equation}
R_0 = \left\{ (x, y, z) : \, \frac{x^2}{a_{1}^2} + \frac{y^2}{b_{1}^2} + \frac{z^2}{c_{1}^2} < 1\right\},
\label{R0}
\end{equation}
where $a_1$, $b_{1}$, and $c_1$ are the semi-axis lengths of the ellipsoid. By definition, the semi-axis length $a_1$ is equal to half of the propagation distance $d_1$, and the semi-axes $b_1$ and $c_1$ are determined by the transmitter–receiver separation $d_0$ and the propagation distance $d_1$, which are respectively given as
\begin{align}
a_1&=\frac{d_1}{2} =\frac{d_0+c \Delta t }{2}, \\
b_1&=c_1=\sqrt{(d_1/2)^2-(d_0/2)^2}.
\end{align} 
Propagation paths in the first arriving region $R_0$ contribute to the major portion of the total received power, since they are shorter than those corresponding to the subsequent arriving regions. Increasing the temporal resolution, i.e., reducing the sampling interval $\Delta t$, shortens the propagation distance $d_1$ and narrows the blue region in Fig.~\ref{F:fresnelzone}, causing the first arriving region to converge to the LoS path. 
Points on the LoS path block EM waves, attenuating the channel gain and leading to obstructed LoS (OLoS) or non-LoS (NLoS)\footnote{Strictly speaking, it is the physical objects represented by the points that interact with EM waves through shadowing, reflection, or scattering. However, since the environment is modeled by point cloud, these interactions can be attributed to the points.}. Points in $R_0$ while deviate from the LoS path act as scattering or reflecting points, redirecting EM waves back to the receiver.

\subsubsection{Subsequent Arriving Region}

After defining the first arriving region, we introduce how to construct the subsequent arriving regions to select points that related to channel gains $\alpha_1, \cdots, \alpha_{(K-1)}$ in the channel model~\eqref{cir}. As shown in Fig.~\ref{F:fresnelzone}, each subsequent arriving region is defined by the region between two adjacent ellipsoids. For arriving region $R_k$, it covers all points that interact with the arriving EM waves arriving within time interval $[t_k,t_{k+1})$, where $t_k = t_0 + k\Delta t$, and due to the finite sampling interval $\Delta t$ of the channel, these EM waves are treated as arriving simultaneously at ToA $t_k$. 

Time interval $[t_k,t_{k+1})$ determines the minimum and maximum propagation distances $d_k$ and $d_{k+1}$, which then determine the sizes of the two adjacent ellipsoids. Thus, the arriving region $R_k$ is defined as 
\begin{equation}
\begin{aligned}
\label{Rk}
R_k = \Bigg\{ (x, y, z) : & \frac{x^2}{a_k^2} + \frac{y^2}{b_k^2} + \frac{z^2}{c_k^2} \geq 1 \text{ and} \\
& \frac{x^2}{a_{k+1}^2} + \frac{y^2}{b_{k+1}^2} + \frac{z^2}{c_{k+1}^2} < 1 \Bigg\}, \quad \forall k \in \mathbb{N} ^+ ,
\end{aligned}
\end{equation}
where $a_k$, $b_k$, and $c_k$ are the semi-axis lengths of the $k$-th ellipsoid.
Same as the first arriving region, the $k$-th ellipsoid's semi-axis length $a_k$ is half of the propagation distance $d_k$, and the semi-axes $b_k$ and $c_k$ are then determined by $d_0$ and $d_k$, which are respectively given as
\begin{align}
\label{eq:dim_k}
a_k&=\frac{d_k}{2} =\frac{d_0+k c\Delta t}{2}, \\
b_k&=c_k=\sqrt{(d_k/2)^2-(d_0/2)^2}.
\end{align}

When the transmitter and the receiver are not located symmetrically along the x-axis, we apply a shift and rotation transformation: First, shift the center of the ellipsoid $\mathbf x_\text{cent} = (\mathbf{x}+\mathbf{x}_\text{T})/2$ back to the origin, and then rotate the $\mathbf{x}$-$ \mathbf{x}_\text{T}$ axis to align with the x-axis. This transformation, applied to both the point cloud and the transceivers, is given by $\mathbf x_\text{p} '= \mathbf{M}_k (\mathbf x_\text{p} - \mathbf x_\text{cent})$, where $\mathbf{M}_k$ is the rotation matrix computed using Rodrigues' rotation formula \cite{rotation}, fully determined by the rotation axis $\hat{\mathbf{\omega}} = \mathbf{v} \times \mathbf{v}_{\text{x}}/\|\mathbf{v} \times \mathbf{v}_{\text{x}}\|$ and the rotation angle $ \theta = \arccos\left(\mathbf{v} \cdot \mathbf{v}_{\text{x}}/\|\mathbf{v} \cdot \mathbf{v}_{\text{x}}\| \right) $, with $\mathbf{v} =\mathbf{x}-\mathbf{x}_\text{T}$ and $\mathbf{v}_{\text{x}} = [1, 0, 0]$. For brevity, we refer readers to find the details in \cite{rotation}.

\begin{figure}[!t]
\centering
\includegraphics[width=0.75\linewidth]{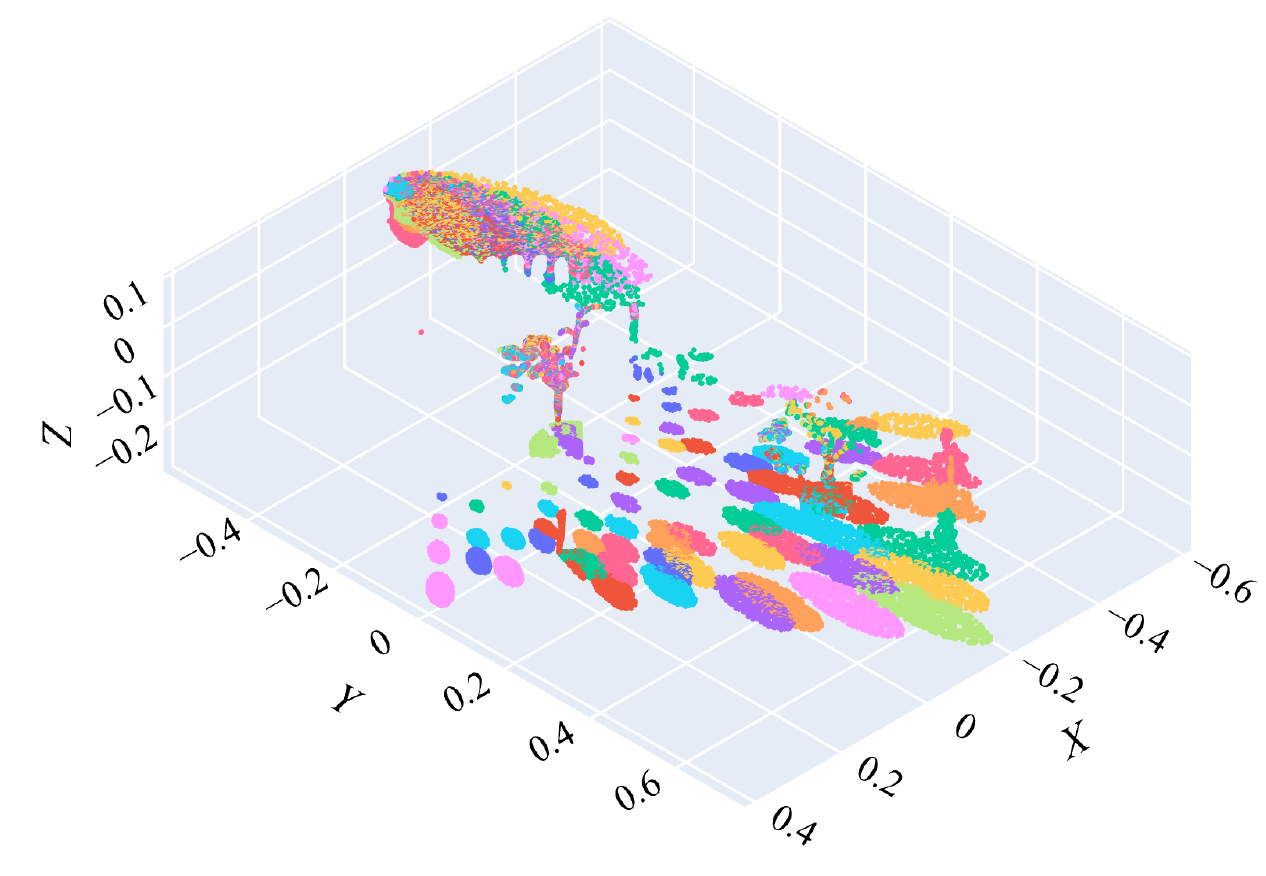}
\caption{Selected subsets of point cloud for different receiver locations at the first ToA.}
\label{ps_los_zhixin}
\end{figure}

In summary, using the defined arriving regions in~\eqref{R0} and \eqref{Rk}, we identify a sequence of subsets of point cloud $\mathcal{P}_k = \left\{ ( \mathbf{p}_i,\mathbf{s}_i)\in \mathcal{P}_\text{env} \vert \mathbf{p}_i \in R_k \right\} $, $k=0,\cdots,K-1$, which are related to the channel gains at the corresponding ToAs. The workflow of Point Selector is summarized in Algorithm~\ref{alg:point_selection2}. 

\begin{remark}
This point selection method establishes a connection between the specific part of the environment and the corresponding channel gain, enabling accurate estimation using only the relevant subset of point cloud. 
Moreover, this method guarantees the inclusion of all points associated with a given ToA, and preserves the original point cloud format to maintain high-precision geometric and auxiliary features.
To visualize the selected subset of point cloud, an example is shown in Fig.~\ref{ps_los_zhixin}, where different colors indicate the subsets of the point cloud corresponding to various receiver locations at ToA $t_0$. 
\end{remark}

\begin{algorithm}
\caption{Point Selector: Selecting Point Cloud $\mathcal{P}_k$ Corresponding to ToA $t_k$} 
\label{alg:point_selection2}
\begin{algorithmic}[1]
\renewcommand{\algorithmicrequire}{ \textbf{Input:}}
\REQUIRE Point cloud $\mathcal{P}_\text{env}$, locations $\mathbf x_\text{T}$ and $\mathbf x$, and time interval $[t_k,t_{k+1})$.

\renewcommand{\algorithmicrequire}{ \textbf{Output:}}
\REQUIRE Selected point cloud $\mathcal{P}_k $.
\renewcommand{\algorithmicif}{\textbf{If}}
\renewcommand{\algorithmicendif}{\textbf{End if}}
\renewcommand{\algorithmicfor}{\textbf{For}}
\renewcommand{\algorithmicendfor}{\textbf{End for}}
\renewcommand{\algorithmicreturn}{\textbf{Return}}

\STATE \textbf{Construct the ellipsoids}:
\STATE Calculate the LOS path length $d_0 = \left\| \mathbf{x} - \mathbf{x}_\text{T} \right\|$.
\STATE Compute the arriving region $R_k$ by \eqref{R0} or \eqref{Rk}.
\STATE \textbf{Define the transformation}:
\STATE Calculate the center of the ellipsoid $\mathbf x_\text{cent} = (\mathbf{x}+\mathbf{x}_\text{T})/2$.
\STATE Determine the rotation matrix $\mathbf{M}_k$ by using Rodrigues’ rotation formula \cite{rotation}.
\STATE \textbf{Select points}:
\FOR{each point $( \mathbf{p}_i,\mathbf{s}_i) \in \mathcal{P}_\text{env}$}
    \STATE Transform its coordinate $\mathbf  x_\text{p} '= \mathbf{M}_k (\mathbf p_i - \mathbf x_\text{cent})$.
    % \STATE Check if the point lies within the bounds of the ellipsoids.
    \IF{$\mathbf  x_\text{p}' \in R_k$}
        \STATE Add point $( \mathbf{p}_i,\mathbf{s}_i)$ to set $\mathcal{P}_k$.
    \ENDIF
\ENDFOR

\end{algorithmic}
\end{algorithm}

\subsection{Neural Channel Gain Estimator}
\begin{figure*}[!t]
\centering
\includegraphics[width=0.8\linewidth]{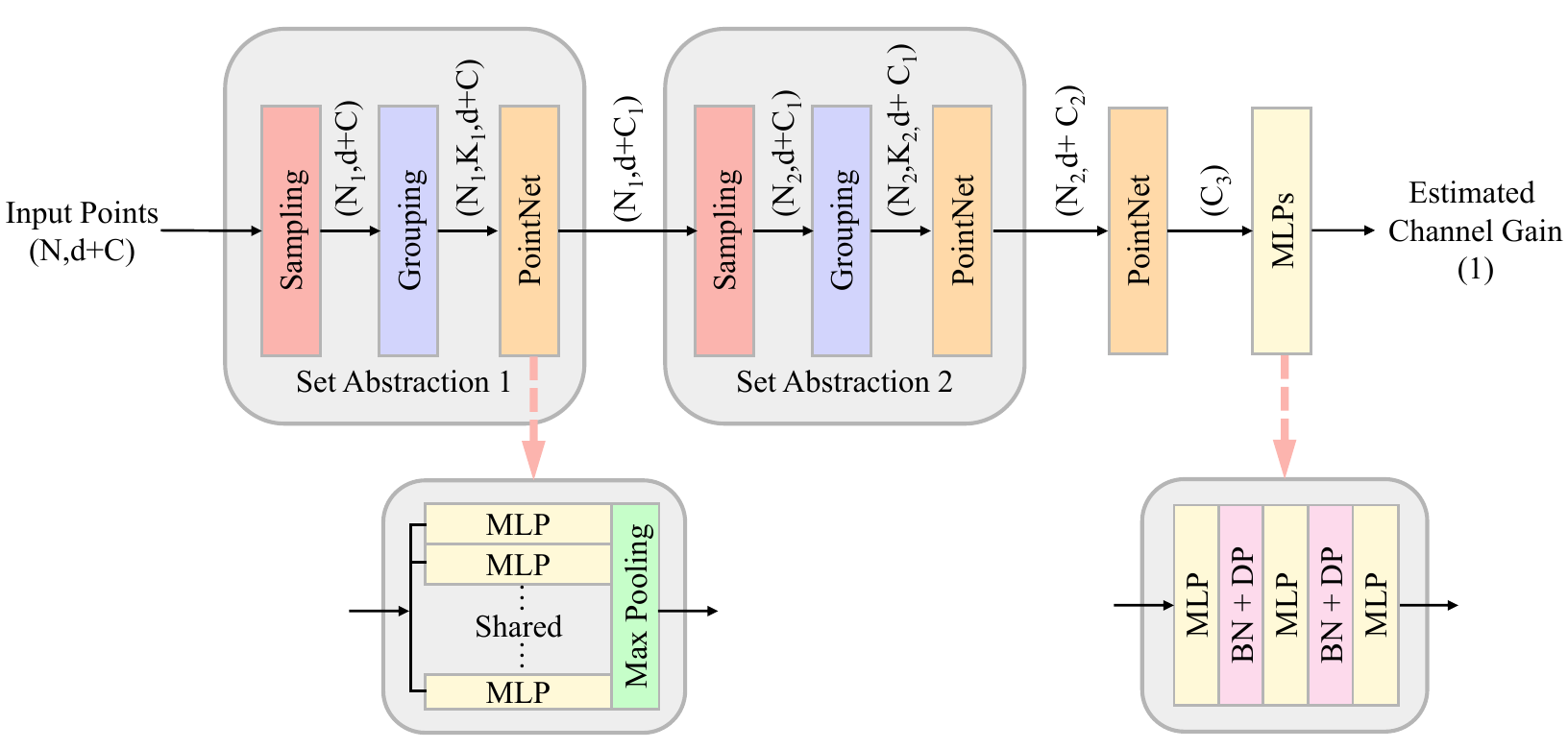}
\caption{Architecture of neural channel gain estimator.}
\label{nn}
\end{figure*}

This subsection introduces Neural Channel Gain Estimator, a data-driven method designed to directly learn the complex mapping from the selected subsets of point cloud to the corresponding channel gains. While channel gain can be estimated using conventional ray-tracing methods, their accuracy is limited by the over-simplified environmental models and inaccurate material parameters.
In contrast, our proposed Neural Channel Gain Estimator learns directly from real-world measurements, thereby bypassing the need for explicit EM modeling and material parameters. By utilizing the detailed geometric information encoded in point clouds, along with carefully designed auxiliary EM wave propagation-related features, the proposed Neural Channel Gain Estimator achieves more accurate channel gain estimation in realistic and complex environments.

As mentioned in the system model, channels are measured at $L$ locations, with $ \{\mathbf h_{1}^o, \mathbf h_{2}^o, \cdots, \mathbf h_{L}^o\}$ being the set of observed PDPs. Each PDP $\mathbf h_{l}^o$ records the channel gain from the first to the last ToA $ [\alpha_{l,0}, \alpha_{l,2}, \cdots, \alpha_{l,K-1}]$, with $\alpha_{l,k}$ being the channel gain at the $l$-th location and the $k$-th ToA. Thus, a total of $S = L \cdot K$ samples are obtained, which are utilized to train the neural estimator to learn the mapping from the selected point cloud to the channel gain. 

We define function $g(\cdot)$ to denote the mapping from the selected point cloud $\mathcal{P}_{k}$ to the channel gain $\alpha_{k}$, i.e.,

\begin{equation}
\alpha_{k} = g(\mathcal{P}_{k}),
\end{equation}
where function $g(\cdot)$ is parameterized by a deep neural network (DNN) $g_{\boldsymbol{\omega}}(\cdot)$ with parameters $\boldsymbol{\omega}$ to be learned. 

\subsubsection{Neural Network Architecture}

As depicted in Fig.~\ref{nn}, the proposed Neural Channel Gain Estimator consists of four key components: the sampling layer, grouping layer, PointNet layer, and multi-layer perceptron (MLP). The network architecture begins with a hierarchical processing structure \cite{ qi2017pointnetplusplus} consisting of two set abstraction levels.
Each set abstraction level aggregates both geometric and propagation-related features from a set of nearby points to create more abstract and higher-dimensional features. Specifically, the first set abstraction level extracts local features within the point cloud, while the second level captures global features. The processed point cloud is then encoded into by a vector using a PointNet, and subsequently mapped to the channel gain estimation by an MLP.

\emph{Input Representation:} The selected point cloud data $\mathcal{P}_k$ is organized into a matrix $\mathbf{X} \in \mathbb{R}^{N \times (d + C)}$ for input, where $N$ is the number of points, $d$ is the coordinate dimension, and $C$ is the features dimension. To ensure a fixed input size, point clouds with more than $N$ points are randomly downsampled, while those with fewer than $N$ points are padded by duplicating existing points.

\emph{First Set Abstraction Level:} Each set abstraction level consists of a sampling layer, a grouping layer, and a PointNet layer.

The sampling and grouping layers detect local geometries while preserving 3D spatial relationships. In the sampling layer, $N_1$ points are sampled using farthest point sampling (FPS), a method that iteratively picks points to maximize the spatial coverage of the point cloud by ensuring each newly selected point is the farthest from the existing sampled set. In the grouping layer, each sampled point identifies its $K_1$ nearest neighbors, forming clusters for local context. 
Importantly, both the sampling and grouping layers are invariant to the reordering of the input points, effectively addressing the challenge of processing unordered points in point cloud. Since point clouds lack inherent order, the network must be invariant to the $N!$ possible permutations of input points.
The output point clouds after this sample and grouping layers are denoted by $\mathbf{X}_1 \in \mathbb{R}^{N_1 \times (d + C)}$ and $\mathbf{X}_2 \in \mathbb{R}^{N_1 \times K_1 \times (d + C)}$, respectively. 

The PointNet \cite{qi2017pointnet} layer then processes the sampled and grouped point cloud. 
Unlike traditional CNN that operate on grid data, PointNet handles unordered sets of points, making it inherently suitable for dealing with the irregular and unevenly spaced point cloud data\cite{qi2017pointnet}. PointNet consists of a MLP and max pooling to encode and aggregate regional features. The MLP is applied to each point to analyzes spatial relationship and incorporate additional features, e.g., color, normal vector, incident angle, and outgoing angle, which influence channel gain. Since the same MLP is shared across all points, this part of network is invariant to the reordering of input points. Max pooling captures the most significant features from each group of points, further ensuring invariance to reordering. The complex interactions between EM waves and different local regions are learned through this process, outputting a point cloud with centroid points representing regional features, i.e. $\mathbf{X}_3 \in \mathbb{R}^{N_1 \times (d + C_1)}$, where $C_1$ is the dimension of the regional feature. 

\emph{Second Set Abstraction Level:} A second set abstraction level is applied to $\mathbf{X}_3 $ to focus on larger spatial regions and extract features from a more global perspective, which generates a smaller subset of point cloud. The point cloud $\mathbf{X}_3 $ after the second sampling, grouping, and PointNet layer are denoted by $\mathbf{X}_4 \in \mathbb{R}^{N_2 \times (d + C_1)}$, $\mathbf{X}_5 \in \mathbb{R}^{N_2 \times K_2 \times (d + C_1)}$ , and $\mathbf{X}_6 \in \mathbb{R}^{N_2 \times (d + C_2)}$, respectively.

\emph{Final Layers:} The final PointNet layer encodes the point cloud $\mathbf{X}_6$ into a vector $\mathbf x_7\in \mathbb{R}^{C_3}$, which is passed through MLPs to estimate the channel gain $\hat{\alpha}_i$. Batch Normalization (BN) and Dropout (DP) layers are inserted between the MLPs to stabilize training and prevent overfitting, respectively.

This architecture effectively utilize the geometric and rich information inherent in the unordered point cloud to accurately estimate channel gain. The loss function and training strategy are discussed in the next subsection.

\subsubsection{Loss Function and Training}

To train the network $g_{\boldsymbol{\omega}}(\cdot)$, the loss function is specially crafted based on the mean squared error (MSE). First, to address training samples with empty point cloud, i.e. $\mathcal{P}_k = \varnothing$, we exclude the corresponding loss, as an empty selected subset of point cloud suggests that there is no propagation path from the transmitter to the receiver. Second, for estimated channel gain $\hat{\alpha}_i$ lower than noise floor value, i.e., $\alpha_{min}$, the channel gain is directly set to the noise floor. Based on the above consideration, the loss function is given as
\begin{equation}
\mathcal{L}(\boldsymbol{\omega})=\frac{1}{S'}
\sum_{i=1}^{S} \mathcal{I}_i
\left\Vert\alpha_i - 
\max\left(\alpha_{min},\hat{\alpha}_i \right) \right\Vert^2,
\end{equation}
where $\hat{\alpha}_i $ is the estimated channel gain, $S'$ is the number of training sample with nonempty point cloud, $S = L \cdot K$ is the size of the entire training set, $ \mathcal{I}_i = 1 $ when the selected point cloud is not empty and $ \mathcal{I}_i = 0 $ otherwise, and $\alpha_i$ is the ground truth channel gain. The loss function is optimized to get the optimal network parameter, i.e., $\boldsymbol{\omega} = \arg\min_{ \boldsymbol{\omega}}\mathcal{L}(\boldsymbol{\omega})$. This optimization is performed using stochastic gradient descent (SGD).

Using a supervised learning approach, the neural network is trained with a novel location-based data splitting strategy to prevent any information leakage from the testing locations during training. Unlike conventional dataset splitting that randomly partitions all $L \cdot K$ samples, we instead divide the $ L $ locations into $ L_1 $ training locations and $ (L- L_1) $ testing locations. This results in $ L_1 \cdot K $ samples used for training and $ (L- L_1) \cdot K $ samples reserved for testing.

\section{Experimental Results and Discussions} 
\label{exp}

In this section, we present comprehensive comparisons of the proposed method with other state-of-the-art approaches. 
Distinct from the majority of existing studies that rely exclusively on synthetic datasets, we conduct experiments based on our own field collected data, which includes both the channel PDP measurements and the environmental point cloud. 

\subsection{Experiments Setup}

Our field data collection was conducted at two sites in the campus of The Chinese University of Hong Kong, Shenzhen (CUHK-Shenzhen). Channel measurements were obtained via a wideband channel sounding system with 600 MHz bandwidth. High resolution point clouds were captured using drone-based photogrammetry, offering detailed 3D representations of the environment. 

\subsubsection{Channel Sounding Setup}

Our wideband channel sounder consists of a Keysight M9484C vector signal generator (VSG) and a Keysight N9032B signal analyzer. On the transmitter side, a sounding sequence with a length of 4096 symbol was generated at a speed of 600 mega-symbol-per-seconds (Msps) and then up-converted to the center frequency of 4.6 GHz in the VSG. The output power from the VSG is 30 dBm. The generated radio frequency (RF) signal was then fed into a horn antenna with a main lobe gain of 12 dBi and half-power beamwidth (HPBW) of 47.8 degree and then radiated to the free space. On the receiver side, the RF signal was received by a omnidirectional antenna with an average gain of -4.5 dBi horizontally at 4.6 GHz. The received signal was down-converted to baseband and then processed by an autocorrelation algorithm \cite{hur2014synchronous} to obtain the PDP. This channel sounding system has a time resolution of $\Delta t =1/B=1.6667$ ns. Both the transmitter and the receiver antenna are vertically polarized. Each measurement was averaged over 10 measurements, thus eliminating the small-scale fading effect.

\subsubsection{Measurement Campaign}
\begin{figure}[tpb]
\centering
\includegraphics[width=1\linewidth]{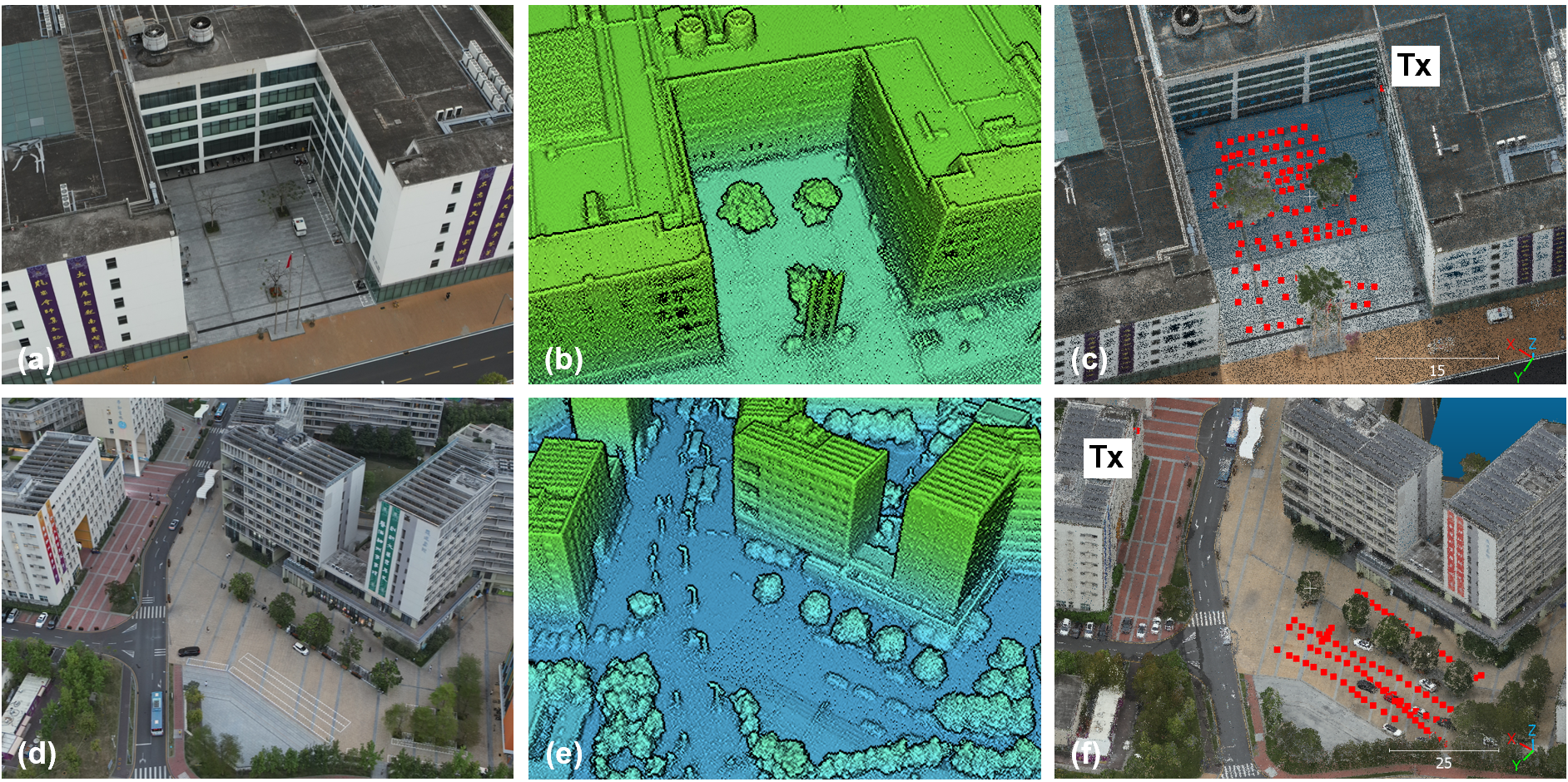}
\caption{Illustration of channel sounding site and the point clouds. First row: the first AoI, located at the lower campus of CUHK-Shenzhen. Second row: the second AoI, located at the upper campus of CUHK-Shenzhen. For each AoI, from left to right. 1: Picture of the environment. 2: Point cloud. 3. Point cloud with object color, and red dots indicating the channel measurement locations.
}
\label{F:scene}
\end{figure}

As shown in Fig.~\ref{F:scene}, the measurement campaign was conducted at two vegetated square sites in campus. A total of 96 locations are measured in the first AoI, and 90 locations are measured in the second AoI. 

The first AoI, shown in Fig.~\ref{F:scene} (a)-(c), is enclosed by the three faces of the building, which are made up of glass windows and concrete wall surfaces. The size of this plaza is $30\times 30$ $\text{m}^2$ and the height of the building is 25 m. The second AoI, shown in Fig.~\ref{F:scene} (d)-(f), is a $30\times 100$ $\text{m}^2$ square with building on two sides of it and with trees in and around the square. In the first AoI, the transmitter antenna is placed at a window at a height of 10.6 m with a downward tilt of 10 degrees. In the second AoI, the transmitter antenna is located on a rooftop terrace at a height of 15 m with a downward tilt of 10 degrees 20 degrees. The receiver is set at a height of 1 m in both sites. 

All these measurement regions resemble the 3GPP urban microcell (UMi) scenario, characterized by surrounding buildings and environmental clutter. Most measured transmitter-receiver links are LoS, while some are obstructed by trees or other obstacles, resulting in OLoS. Additionally, a small portion of the measured links are NLoS, where no LoS path exists, where the received power mainly arises from reflection, scattering, and diffraction.

\subsubsection{Point Cloud Acquisition}
To obtain the point cloud of an AoI, an unmanned aerial vehicle (UAV) equipped with a camera captures multi-view images from above the AoI. These images are then processed using 3D environment reconstruction software that applies photogrammetry to generate a 3D model of the environment. 

Specifically, we use a DJI M300 RTK drone equipped with a Zenmuse P1 camera to conduct the 3D aerial survey at CUHK–Shenzhen. The UAV followed a pre-planned route, generated by a route planning application based on the desired reconstruction accuracy, capturing images from multiple angles. The images were then processed by DJI Terra software through the photogrammetry method to generate a mesh model, and then processed by CloudCompare software to generate a point cloud with detailed features, including normal vectors and color. 

The point cloud generated for the CUHK-Shenzhen campus consists of approximately 58.9 million points. The reconstruction root mean square error (RMSE) reported by DJI Terra software was 2.5 cm, indicating a very small deviation between reconstructed points and true locations. This error varied across different objects, with transparent objects like glass showing larger errors compared to opaque surfaces. Accurate reconstruction of transparent objects remains a challenge in 3D environment reconstruction\cite{Qian_2016_CVPR}.

\subsection{Point Selector Validation} \label{sec:psp}

In this subsection, we demonstrate the proposed Point Selector's capabilities in selecting points correlated to a given ToA and compare its performance with other methods. Identifying the points precisely is critical for effectively utilizing the environmental information and improving construction accuracy. As noted in the related works in the introduction, only ray-tracing methods are capable of utilizing 3D environmental information. Therefore, we compare Point Selector with two representative methods from this category. 

\begin{itemize}
   \item \emph{Wireless Insite Ray-Tracing} \cite{_wireless}: As one of the most widely used and recognized ray-tracing tools in wireless channel prediction, many radio map construction studies employ this software to generate ground truth datasets. The used environment information is a mesh-based model and the material properties are set according to the ITU recommendation \cite{ITU_RP2040}. 
   \item \emph{Point Cloud Ray-Tracing} \cite{jarvelainen2016indoor,koivumaki2021impacts,virk2018site}: These studies adapted ray-tracing for point cloud data. Similar to other ray-tracing methods, this approach first simplifies the environment, and then uses EM wave propagation models and permittivity parameters to calculate propagation loss. Flattened surface are constructed from point cloud based on the article's setting by choosing 8 closest points to fit a plane, and permittivity are obtained using the measured PDP.
  
\end{itemize}

\begin{figure*}
    \centering
    \includegraphics[width=1\linewidth]{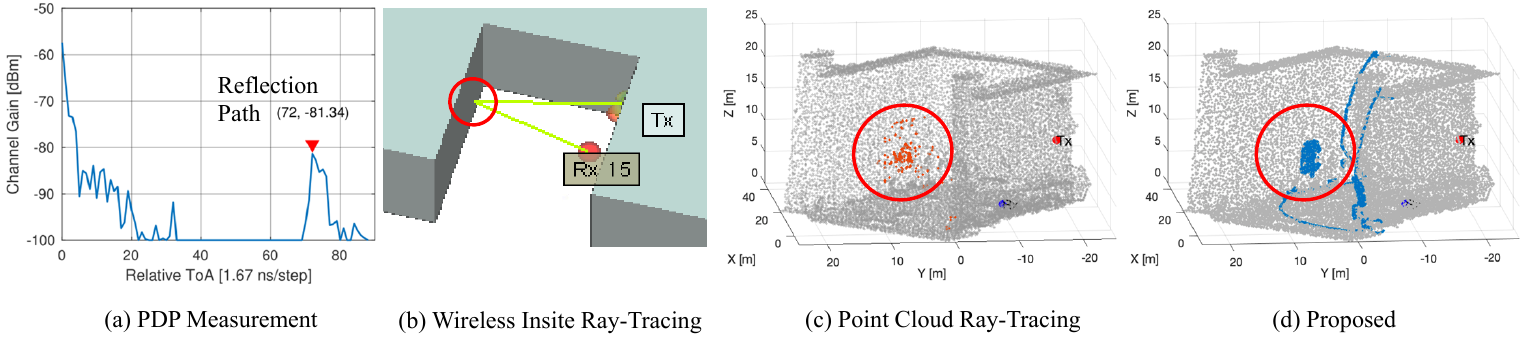}
    \caption{
    A comparison example of identifying the reflection points at the 15th receiver location in the 1st AoI.}
    \label{f_ray_tracing}
\end{figure*}

For reflection points identification, a typical example is shown in Fig.~\ref{f_ray_tracing}. As shown in the measured PDP in Fig.~\ref{f_ray_tracing}(a), there is a significant peak at ToA step 72, corresponding to a strong reflection path. Wireless Insite, shown in Fig.~\ref{f_ray_tracing}(b), correctly identifies the reflection point but detects only a single point. The Point Cloud Ray-Tracing method, shown in Fig.~\ref{f_ray_tracing}(c), is capable of detecting a cluster of points where the reflection occurs. In contrast, our method, shown in Fig.~\ref{f_ray_tracing}(d), identifies a more complete subset of point cloud. A comparison between Fig.~\ref{f_ray_tracing}(c) and (d) reveals that the Point Cloud Ray-Tracing method tends to include points that are distant from the actual reflection location, leading to inaccurate reflection identification. 
In contrast, the proposed Point Selector accurately identifies points relevant to the specified ToA and showing that the region around the reflection point also contributes to the reflection. This capability enables our approach to more accurately identify reflection points and to estimate the channel gain with higher accuracy.

\begin{figure*}
    \centering
    \includegraphics[width=0.75\linewidth]{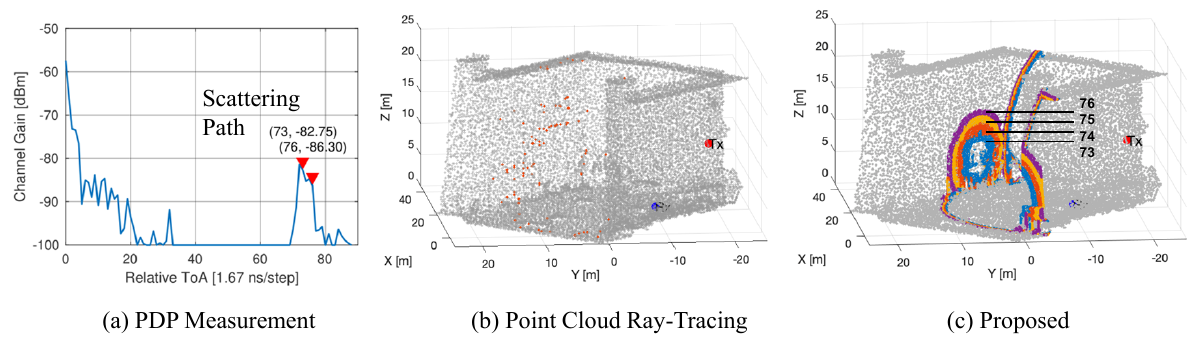}
    \caption{
    A comparison example of identifying the diffuse scattering objects at the 15th receiver location in the 1st AoI. }
    \label{f_ray_tracing_ds}
\end{figure*}

For diffuse scattering points identification, Point Selector also demonstrates superior performance as shown in Fig.~\ref{f_ray_tracing_ds}. Following the reflection path at ToA step 72, there exist several subsequent diffuse scattering paths, shown in Fig.~\ref{f_ray_tracing_ds}(a), which is often disregarded in conventional methods. 
Wireless Insite uses the Lambertian scattering model \cite{4052607}, which assumes that scattering occurs across the entire wall. It divides the wall into sections for calculation, without identifying specific points, and is therefore excluded from comparison. 
Point Cloud Ray-Tracing, shown in Fig.~\ref{f_ray_tracing_ds}(b), identifies a limited number of discrete scattering points on the wall and ground, but is unable to find a complete region. In contrast, our method, shown in Fig.~\ref{f_ray_tracing_ds}(c), identifies more complete subsets of point cloud and reveals concentric ring-shaped regions corresponding to distinct ToAs, where the inner ring for early-arriving paths and successive outer rings for later-arriving paths. This concentric ring-shaped scattering observations indicate that scattering does not occur uniformly over a surface but instead concentrates within rings surrounding the reflection area.

Therefore, Point Selector offers an accurate and comprehensive identification of points based on the ToA, overcoming the limitations of conventional ray-tracing methods, which fail to effectively leverage point cloud data. Its effectiveness is grounded in the theoretical framework presented in the previous section, with consistent results observed across different locations and ToAs. Furthermore, its efficacy is validated by the results in downstream CKM construction.
 
\subsection{CKM Construction Results}

In this subsection, we compare the PDP-based CKM construction performance of the proposed methods with other benchmarks. As outlined in the previous subsection, only ray-tracing methods, among those discussed, are capable of estimating the PDP. Consequently, our comparisons will focus on the two previously mentioned benchmarks.

We divided the dataset into two segments: training and testing. Specifically, 80\% of the locations were allocated for training the model, while the remaining 20\% were designated for testing the model's accuracy.

\subsubsection{Metric}
To quantitatively assess the estimation accuracy of the model, the RMSE metric is employed, where the RMSE is computed as $\text{RMSE}(\mathbf h,\mathbf h^*) = \sqrt{\|  \mathbf h - \mathbf h^*\|_2^2/N}$. This metric effectively captures the deviation of the estimated PDP, $\mathbf h^*$, from the actual PDP, $\mathbf h$, offering a direct measure of estimation error in decibels, thus enabling a straightforward assessment of model precision.

\subsubsection{Construction Results}
We evaluate the performance of the proposed method with other ray-tracing benchmarks by comparing the estimated and measured PDPs at three representative locations. These comparisons demonstrate the method's effectiveness in capturing the main peaks and overall trends of the measured PDPs. In addition to these specific examples, we present a comprehensive evaluation using the RMSE metric across different methods, highlighting the superior accuracy and consistency of our proposed approach.

\begin{figure*}[!t]
    \centering
    \includegraphics[width=1\linewidth]{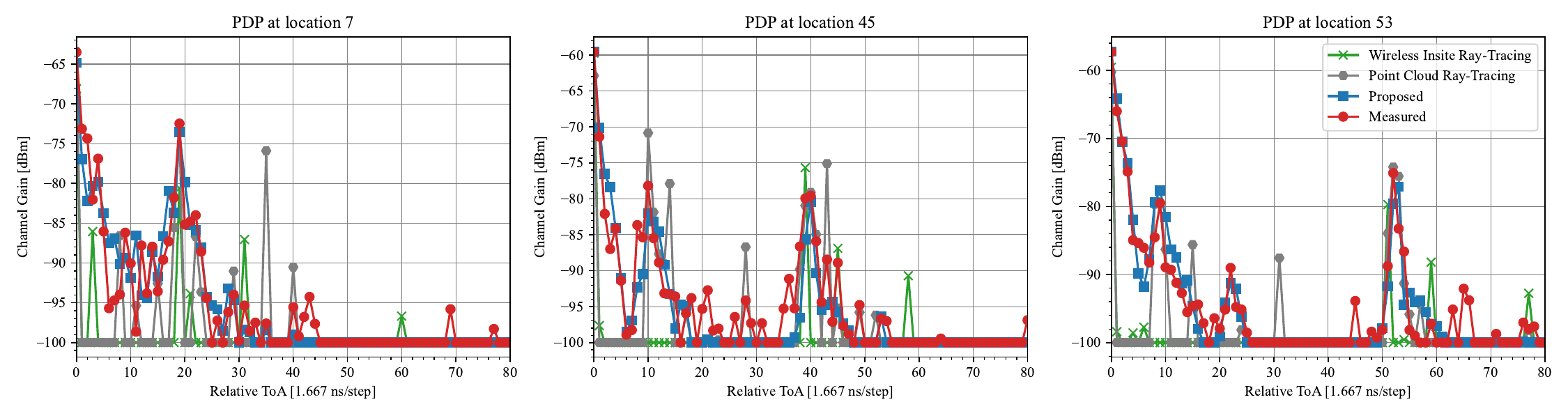}
    \caption{Examples of PDP constructions with proposed method and various benchmark in the 1st AoI.}
    \label{F:cir_measure}
\end{figure*}

We present three examples that illustrate the comparison between the estimated and measured PDP in Fig. \ref{F:cir_measure}. Each figure plots the channel gain versus ToA steps, with a time resolution of 1.667 ns per step. For easier visualization, these figures use relative ToA, i.e., the arrival time of the first path is set to zero. In these examples, the estimated PDP of the proposed method closely follows the measured PDP with a high degree of accuracy, capturing the main peaks and the overall trend. Although some the estimated values show slight discrepancies in the height of the peaks, the model effectively identifies the significant multipath components and the temporal alignment is precise. To sum up, the estimation not only aligns well in terms of the timing of the peaks but also in their magnitude, indicating a robust estimation capability of the model. 

The PDP estimations from other ray-tracing benchmarks are also shown is Fig.~\ref{F:cir_measure}. While conventional ray-tracing methods succeed in identifying reflection path in the measured PDP (peaks in the measured PDP), they also introduce error paths where the measured PDP lacks any corresponding peaks. These inaccuracies can significantly reduce estimation accuracy. Furthermore, although these methods may predict reflection ToA with relative accuracy, they fail to estimate the channel gain accurately. These discrepancies in channel gain estimation are notably significant, with errors ranging from 5-10 dB, as seen at location 7 at ToA step 3 in Fig.~\ref{F:cir_measure}. Additionally, they have difficulty identifying scattering paths following a reflection path and fail to detect reflection paths with relative smaller channel gain. As a result, these methods can lead to substantial inaccuracies in the overall channel PDP estimation.

\begin{figure}
    \centering
    \includegraphics[width=0.9\linewidth]{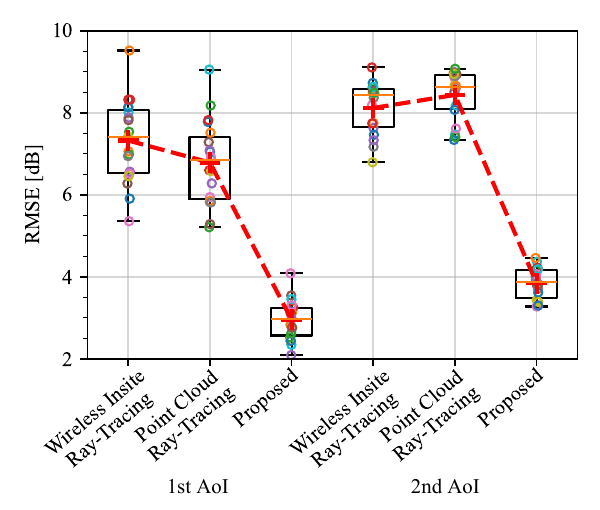}
    \caption{RMSE comparison of the proposed method with other benchmarks in the two AoIs. }
    \label{F_box_error}
\end{figure}

The overall RMSE values for different methods are presented in box plot shown in Fig.~\ref{F_box_error}, which shows the RMSE for the testing locations at the two AoIs. From the figure, it is evident that our proposed method achieves the lowest PDP estimation RMSE error at both AOIs, with an average value of 2.95 dB and 3.84 dB, respectively. In comparison, Wireless Insite Ray-Tracing method shows a higher RMSE, with an error of 7.32 dB and 8.11 dB, while the Point Cloud Ray-Tracing method's RMSE is 6.79 dB and 8.43 dB. This demonstrates that our method provides a significant improvement, with a reduction of approximately 4 dB in error compared to both methods in both AoIs. Furthermore, our proposed method not only shows a smaller mean error but also shows a lower variance in error distribution, emphasizing its improved accuracy and consistency. 

\subsection{Radio Maps Construction Results}

In this subsection, we further evaluate the proposed method's capabilities in CKM construction of received power values, i.e. radio map. In addition to the previously mentioned ray-tracing methods, we also include an AI-based method and a interpolation method for comparison. The settings for these newly included method are described as follows:

\begin{itemize}
    \item \emph{RadioUNet Neural Network} \cite{levie2021radiounet}: RadioUNet employs an image-to-image deep learning approach for estimating the radio map, where the input images consist of a 2D building map and an image indicating the transmitter location. The output is the constructed radio map for a given AoI. Since the pixels values of the output image range from 0 to 255, a post-processing step is performed to align the estimation with the measured values using the training set.
    
    \item \emph{Kriging Interpolation} \cite{chowdappa2018distributed}: Originated from the geostatistical interpolation, this methods is now widely used to construct the radio map. The Regression Kriging (RK) is implemented for comparison and the semivariance model adopts the exponential model.

\end{itemize}

\begin{figure*}[!ht]
    \centering
    
    \subfloat[Proposed\label{fig:sub1}]{
        \includegraphics[width=0.25\linewidth]{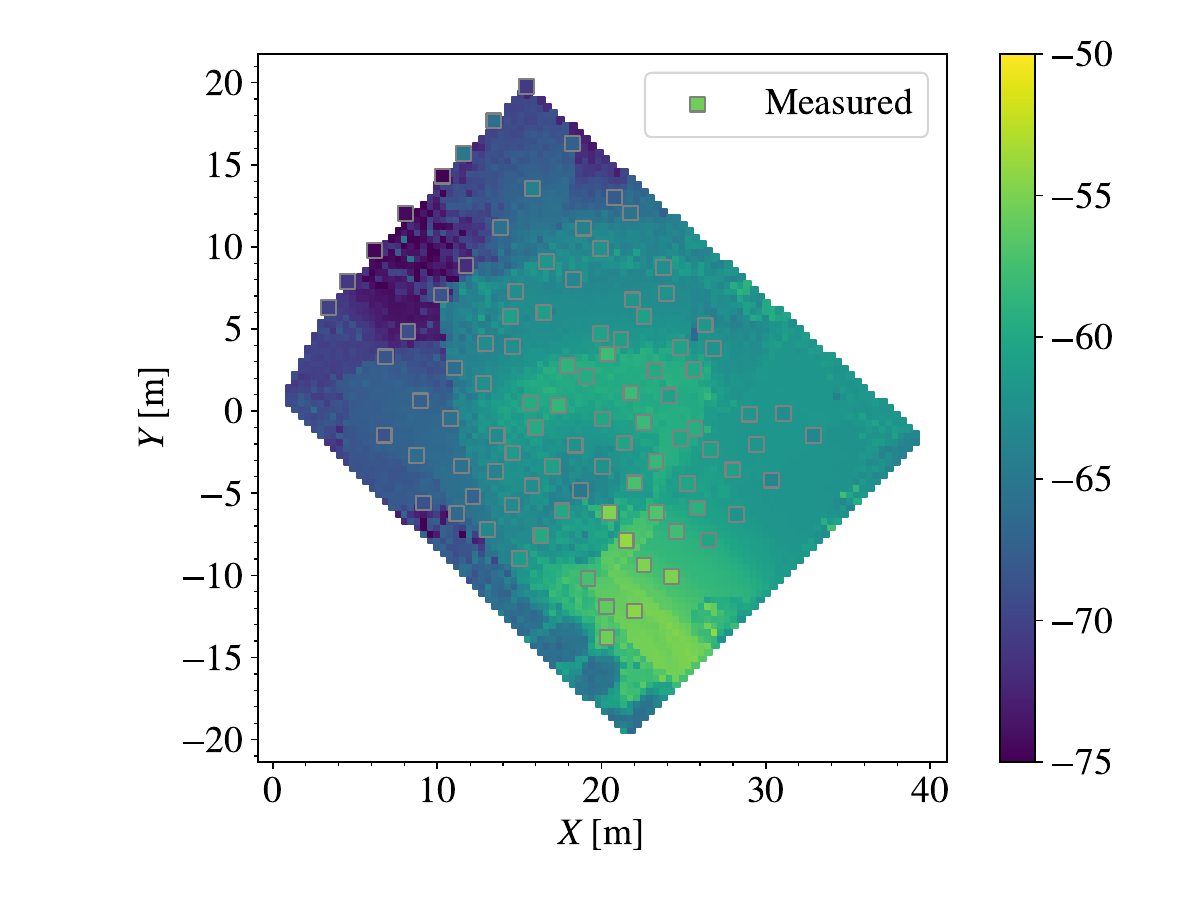}
    }
    % \hfill
    \subfloat[Wireless Insite Ray-Tracing\label{fig:sub2}]{
        \includegraphics[width=0.25\linewidth]{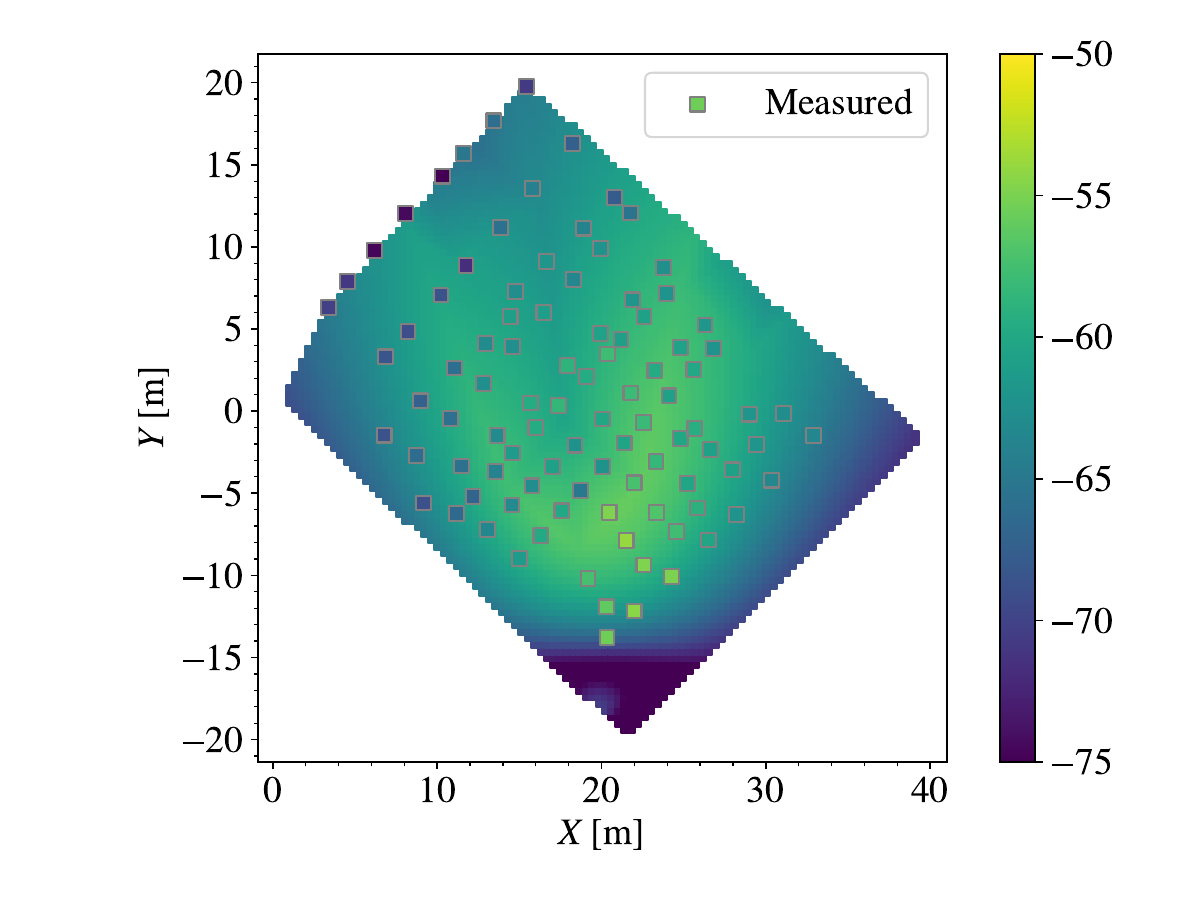}
    }
    % \hfill
    \subfloat[RadioUNet\label{fig:sub3}]{
        \includegraphics[width=0.25\linewidth]{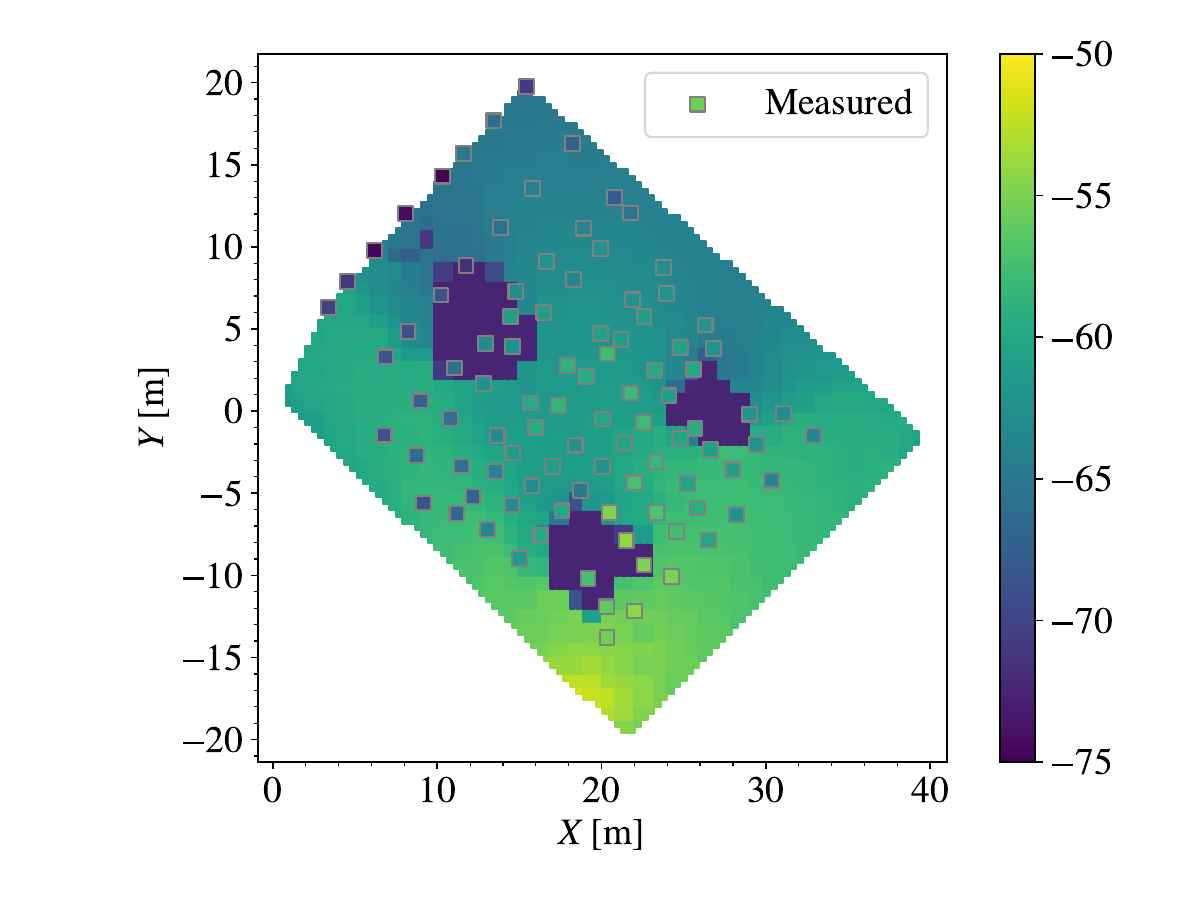}
    }
    \subfloat[Kriging\label{fig:sub4}]{
        \includegraphics[width=0.25\linewidth]{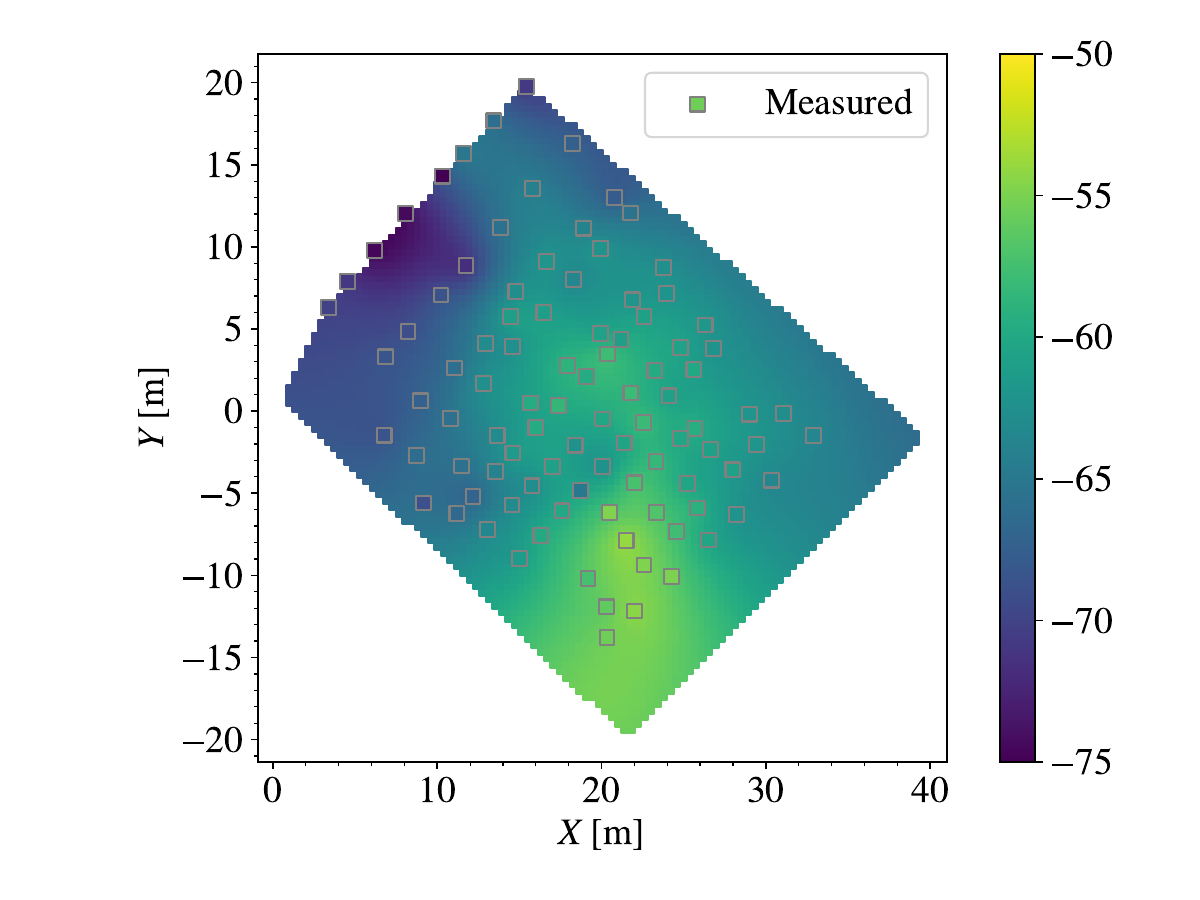}
    }
    % \hfill
    % \subfloat[Scattering Objects\label{fig:sub4}]{
    %     \includegraphics[width=0.24\linewidth]{fig/ds.png}
    % }
    
    \caption{A radio map of received power values constructed from sampled points at a fixed transmitter position in the 1st AoI.}
    \label{F:RM_result_zx}
\end{figure*}

\begin{figure*}[!ht]
    \centering
    
    \subfloat[Proposed\label{fig2:sub1}]{
        \includegraphics[width=0.25\linewidth]{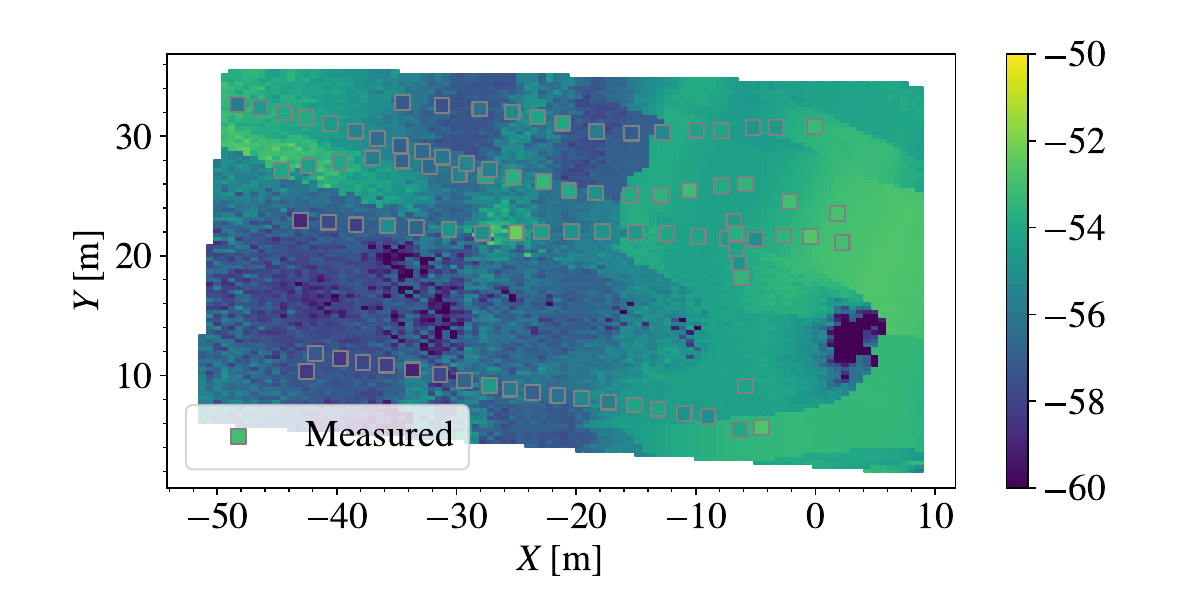}
    }
    % \hfill
    \subfloat[Wireless Insite Ray-Tracing\label{fig2:sub2}]{
        \includegraphics[width=0.25\linewidth]{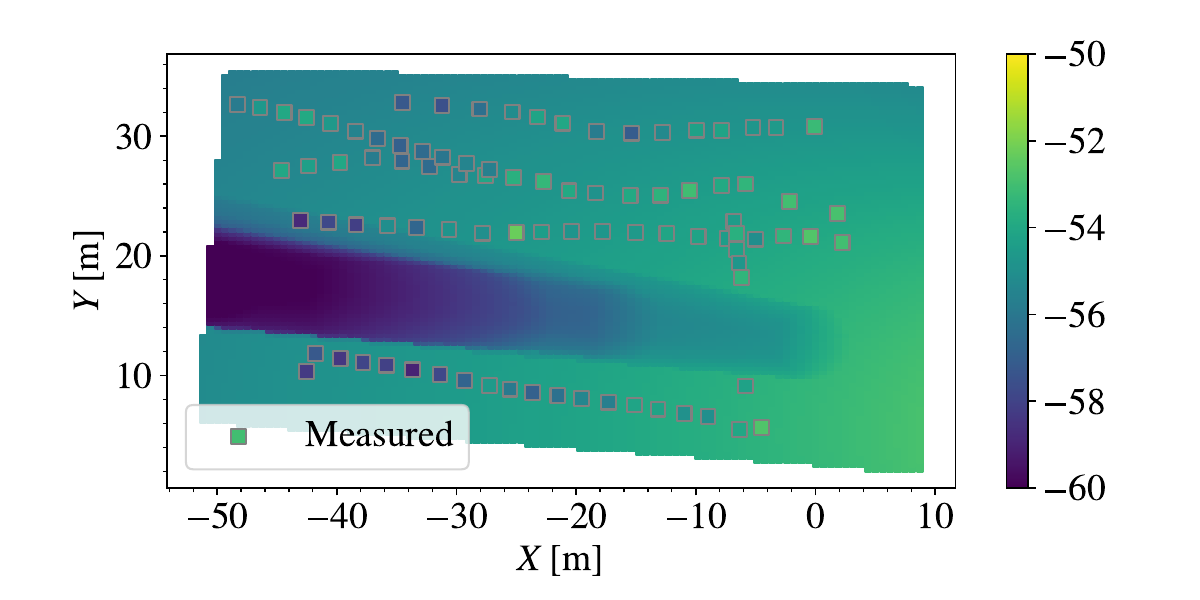}
    }
    % \hfill
    \subfloat[RadioUNet\label{fig2:sub3}]{
        \includegraphics[width=0.25\linewidth]{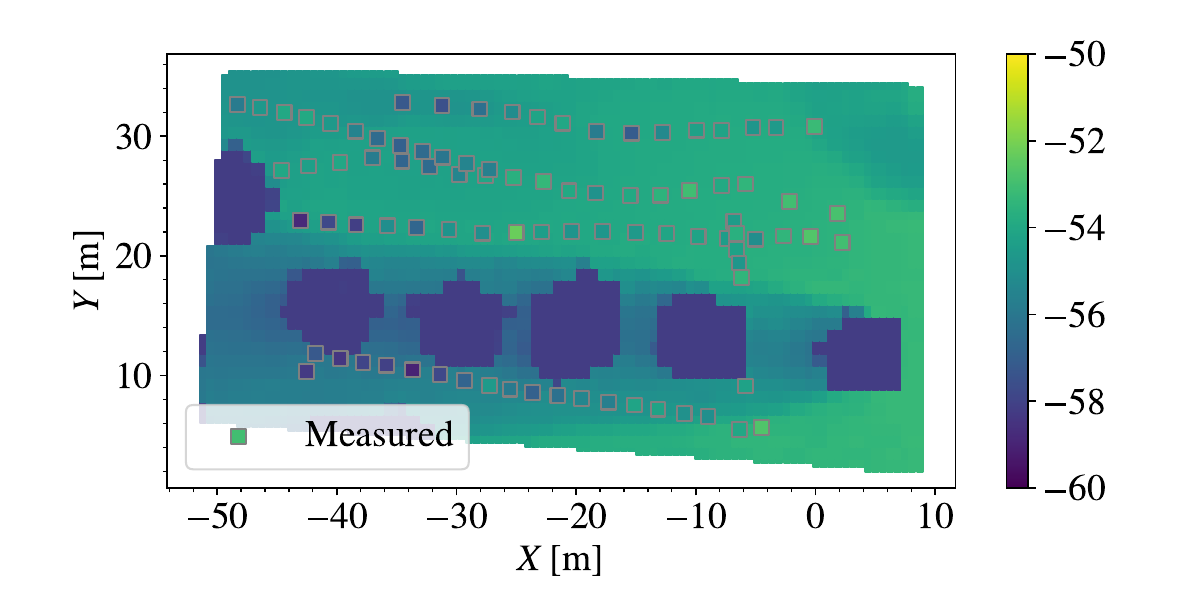}
    }
    \subfloat[Kriging\label{fig2:sub4}]{
        \includegraphics[width=0.25\linewidth]{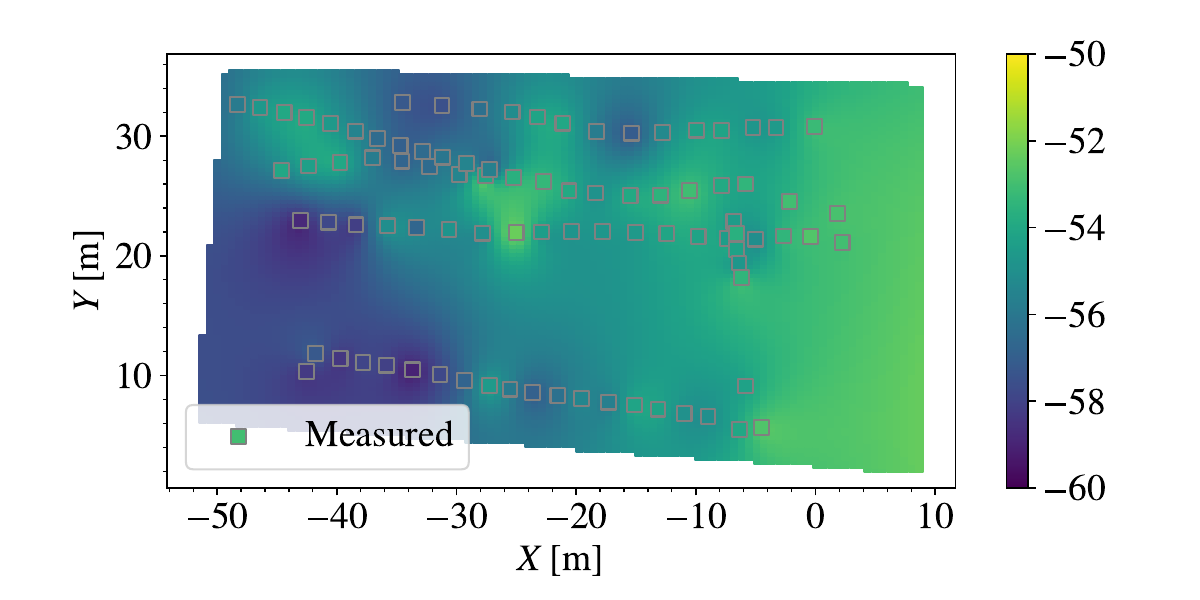}
    }
    % \hfill
    % \subfloat[Scattering Objects\label{fig:sub4}]{
    %     \includegraphics[width=0.24\linewidth]{fig/ds.png}
    % }
    
    \caption{A radio map of received power values constructed from sampled points at a fixed transmitter position in the 2nd AoI. }
    \label{F:RM_result_sy}
\end{figure*}

\begin{figure}
    \centering
    \includegraphics[width=0.8\linewidth]{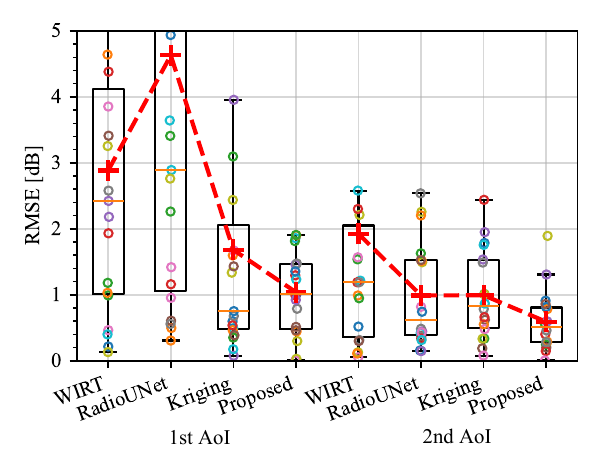}
    \caption{RMSE comparison of the proposed method with other benchmarks in the two AoIs. }
    \label{F_box_error_2}
\end{figure}

The comparative results of radio map construction in the two AoIs are shown in Fig.~\ref{F:RM_result_zx} and \ref{F:RM_result_sy}, where the measured received power values are marked with gray squares in theses figures. To evaluate the quality of the constructed radio map, we examine the similarity between the measured values and the surrounding estimated values in the radio map. A smaller color difference between the measured points and their surrounding areas indicates higher construction accuracy. 

All four methods successfully capture the overall trend of the radio map (free-space propagation loss), where locations closer to the transmitter receive stronger signals, and the signal strength decreases as the distance increases. However, in regions of the radio map where irregular variations arise from more complex phenomena, i.e., shadowing, reflection, and scattering, the performance of the methods differs significantly. These differences reveal the distinctive characteristics and varying accuracy of each approach.

As shown in Fig.~\ref{F:RM_result_zx}(a), the proposed method provides a highly accurate radio map construction result and captures fine‐grained details with remarkable precision. For example, the dark region in the upper‐left and the bright region in the lower‐right clearly reflect the method’s ability to learn environment-specific propagation characteristics from the measurement data. It also accurately reflects the shadowing phenomenon: there are three trees in the environment, but since the two closer trees are not very dense and the antenna is relatively high, their impact on signal attenuation is not severe. In contrast, the more distant tree exerts a significant blocking effect on measurements, leading to more obvious attenuation. These difference in the environment, which depend on three‐dimensional environmental information, are well utilized by this method. In other words, the learned model is not a simple, uniform propagation model; rather, it is highly correlated with the actual point cloud environment, enabling it to derive valuable insights from the relationship between the environment and the measurement data. 

As shown in Fig.~\ref{F:RM_result_zx}(b), Wireless Insite Ray‐Tracing provides a reasonably accurate estimation of the large‐scale attenuation. However, noticeable differences appear in regions with complex multipath or obstacles, which is not fully captured by ray‐tracing, i.e., in the top‐left region. The resulting radio map appears smoother compared to the measured points. 

As shown in Fig.~\ref{F:RM_result_zx}(c), the AI-based RadioUNet successfully captures both free-space propagation loss and shadowing effects, clearly demonstrating the shadowing effects of the three trees in the environment. However, its predictions still show notable deviations from actual measurements. For instance, the signal attenuation caused by the two closer trees is minimal in the measured data due to their sparse foliage, yet RadioUNet overestimates their shadowing effects due to its inability to utilize 3D environmental information. Moreover, the scattering effects introduced by trees are highly complex and difficult to model accurately without using a joint model- and data-driven approach. The model also tends to overestimate signal strength near the transmitter, further contributing to its mismatch with real-world measurement results.
 
As shown in Fig.~\ref{F:RM_result_zx}(d), the Kriging interpolation method achieves relatively good accuracy in constructing the radio map. It effectively captures the global trends in this AoI, with weaker signals toward the top and stronger signals near the bottom. However, due to it lack of physical propagation modeling and inability to incorporate environmental information, the resulting radio map is overly smooth, especially in cases with sparse measurements. Regions with abrupt variations, e.g., those caused by trees or other obstacles leading to rapid signal drops, are not well captured. Moreover, the lack of sufficient measured data in edge regions, e.g., lower-left corner, further degrades accuracy, reflecting a common limitation of spatial interpolation methods. The results in Fig.~\ref{F:RM_result_sy} are similar to those in Fig.~\ref{F:RM_result_zx}, and thus, further explanation for the second AoI is omitted.

The RMSE values for radio map construction using different methods are shown in the box plot in Fig.~\ref{F_box_error_2}. Our proposed method achieves the lowest RMSE, with values of 1.04 dB and 0.59 dB for the two AoIs, respectively. In comparison, Wireless Insite Ray-Tracing shows higher RMSE values of 2.88 dB and 1.92 dB, while the Kriging method's RMSE is 1.68 dB and 0.99 dB. For RadioUNet, which outputs pixel values ranging from 0 to 255, the conversion to received power introduces substantial errors. This issue is especially pronounced in tree-covered areas, where the model predicts no signal reception within obstacles, making direct numerical comparison less meaningful. These results demonstrates that our method significantly outperforms both Wireless Insite and Kriging, offering more accurate and consistent results. By leveraging real-world environmental information, the proposed method captures the intricate relationship between the environment and the wireless channel. Combining 3D environmental information with measured values, it achieves higher estimation accuracy in complex, real-world conditions, outperforming traditional ray-tracing, interpolation, and AI-based methods, and providing a more accurate reconstruction of complex propagation phenomena.

\subsection{Influence of Point Cloud Density on CKM Reconstruction Accuracy}

\begin{figure*}
    \centering
    \includegraphics[width=1\linewidth]{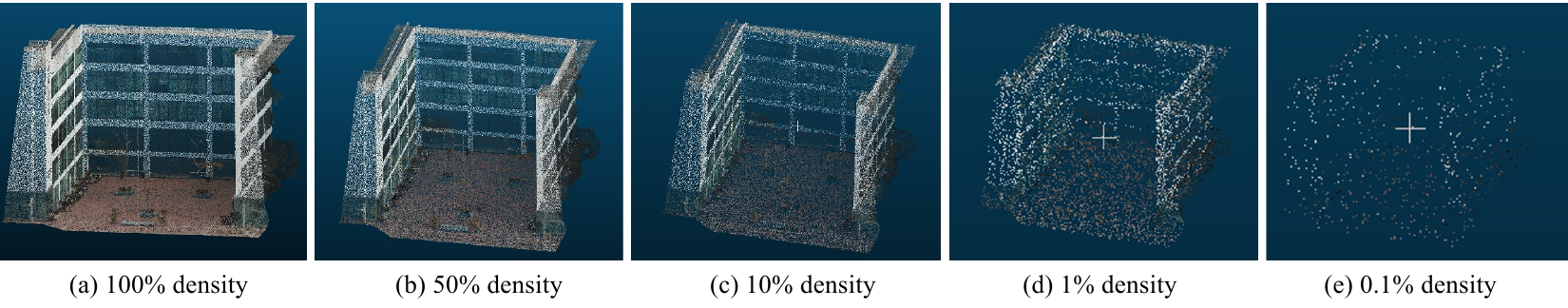}
    \caption{Illustration of the point cloud with different density at the 1st AoI. }
    \label{pc_ds}
\end{figure*}

\begin{figure}
    \centering
    \includegraphics[width=0.8\linewidth]{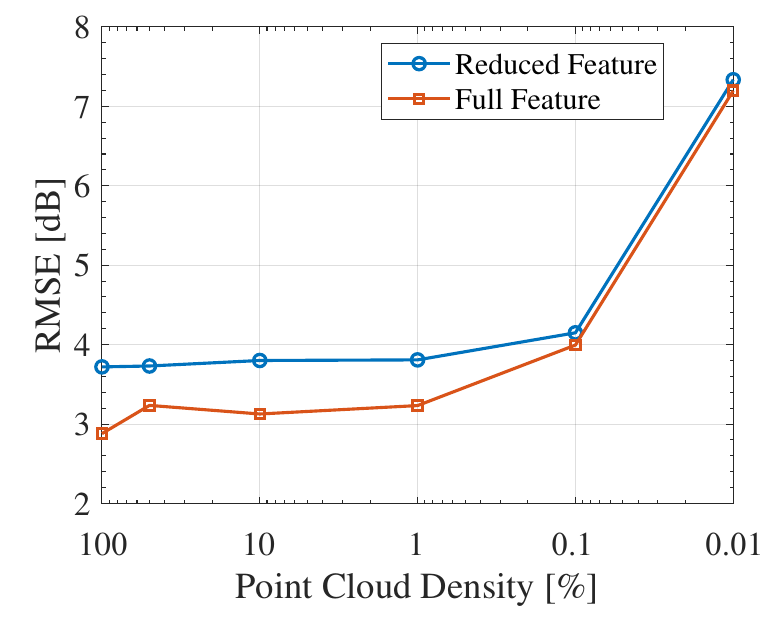}
    \caption{Comparison of the CKM construction accuracy under different point cloud density. }
    \label{pc_sd_o}
\end{figure}

To further analyze the impact of point cloud on CKM construction, we investigate the relationship between CKM accuracy and point cloud density in this subsection, as the point cloud serves as a critical input for the proposed methods. As shown in Fig.~\ref{pc_ds}, this experiment is conducted using down-sampled point cloud data from the first AoI. The original point cloud contains 2 million points (representing 100\% density), and we perform down-sampling at different ratios to assess performance under varying densities.

As shown in Fig.~\ref{pc_sd_o}, the results are presented in two cases: the first (red line) includes the full corresponding features $\mathbf s_i$ for each point, including color, normal vector, distance, and incident and outgoing angle. The second case (blue line) includes only the coordinates and color information for each point. Notably, the blue line consistently lies above the red line across different densities, indicating that providing more auxiliary information greatly contributes to more accurate CKM estimation. 

We also observe that as the density of the point cloud decreases, the estimation error initially grows gradually. However, once the point cloud density is insufficient to capture the overall shape of the scene, i.e., lower than 1\% percent, the error increases more rapidly. Significant performance degradation is particularly evident at 1\% and 0.1\% densities. We attribute this is to the relatively regular structure of the environment in this experiment, which still allow acceptable CKM construction accuracy even at lower point cloud densities. In this environment, the dominant paths are mainly LoS path and reflections path from flat surfaces, which can be captured with lower point density. To validate this hypothesis, further experiments in more complex environments are needed to better understand how structural complexity influences the required point cloud density.

Overall, the results demonstrate that reduced point cloud density leads to lower construction accuracy, emphasizing the importance of high-precision point cloud data for accurate CKM construction.

\section{Conclusions}
\label{conclusion}

In this paper, we introduced a joint model- and data-driven method for CKM construction by leveraging 3D point cloud environmental data and location-tagged channel information. The approach consists of two key modules: Point Selector and Neural Channel Gain Estimator. The Point Selector effectively utilizes the 3D point cloud environmental information and identifies subsets of point cloud that related to channel gains for the given ToAs. The Neural Channel Gain Estimator then maps these identified subsets of point cloud to channel gains, utilizing the geometric and auxiliary information inherent in the point cloud. To validate this method, we conducted field data collection for both the channel PDP measurements and the environmental point cloud across two campus AoIs. Experimental results shows that the proposed methods outperforms conventional ray-tracing methods and SOTA AI-based methods. By integrating detailed environmental information into the channel estimation process, our approach addresses a critical gap in CKM construction and channel modeling, offering a promising solution for next-generation wireless communication systems, where accurate channel knowledge is urgently needed.

\bibliographystyle{IEEEtran}

\bibliography{IEEEabrv,reference}

\end{document}